\documentclass[aps,amssymb,amsmath,preprint,superscriptaddress,nofootinbib,tightenlines]{revtex4-1}
\usepackage{graphicx,xcolor,hyperref,setspace,physics,cancel,tensor,longtable}
\pdfoutput=1

\newcommand{\fdiagram}[2][20ex]{\begin{minipage}{#1}\includegraphics[width=#1]{#2}\end{minipage}}

\begin{document}
\title{The UV sensitivity of the Higgs potential\\
in Gauge-Higgs Unification}
\author{Atsuyuki Yamada}
\affiliation{Department of Physics, Nagoya University,
Nagoya 464-8602, Japan}

\begin{abstract}
In this paper, we discuss the UV sensitivity of the
Higgs effective potential in a Gauge-Higgs Unification (GHU)
model.
We consider an $SU(\mathcal N)$ GHU on $\mathbf M^4\times S^1$
spacetime with a massless Dirac fermion.
In this model, we evaluate the four-Fermi diagrams at
the two-loop level and find them to be logarithmically
divergent in
the dimensional regularization scheme.
Moreover, we confirm that their counter terms contribute
to the Higgs effective potential at the four-loop level.
This result means that the Higgs effective potential
in the GHU depends on UV theories as well as in other
non-renormalizable theories.
\end{abstract}
\maketitle

\section{Introduction}
The standard model (SM) of particle physics is a Yang-Mills
theory symmetric under $SU(3)_c\times SU(2)_L\times U(1)_Y$
gauge transformations.
In the SM, the gauge symmetry is spontaneously broken via
the Higgs mechanism, which is caused by the nonzero
vacuum expectation value (VEV) of the Higgs boson.
As a result, physical quantities such as the mass of a particle
include the Higgs VEV.
By measuring the physical parameters including the Higgs VEV,
the SM is confirmed to be consistent with phenomena at Large
Hadron Collider \cite{ATLAS:2018doi,CMS:2018lkl}.

While the phenomenology below the electroweak (EW) scale is
understandable by the SM, it is concerned that the SM has
difficulty explaining the scale hierarchy between the
EW and UV theories such as the grand unified theory (GUT) or
quantum gravity because of the dangerous quadratic
divergences derived from the Higgs boson.
In a model with supersymmetry (SUSY)
\cite{Veltman:1980mj,Davier:1979hr,Dimopoulos:1981au,
Witten:1981nf,Dine:1981za},
there is a superpartner
for each particle to cancel the quadratic
divergences consequently.
Instead of the SUSY scenarios, we can invoke gauge symmetry in
a non-SUSY theory defined on extra-dimensional spacetime
to protect the Higgs mass term.
This higher-dimensional gauge theory is called the gauge-Higgs
unification (GHU)
\cite{Fairlie:1979zy,Manton:1979kb,Forgacs:1979zs,
Hosotani:1983xw,Hosotani:1983vn,Hosotani:1988bm}.

In the GHU models, the Higgs bosons are identified with
the Yang-Mills Aharonov-Bohm (AB) phases.
Therefore, the Higgs boson has no potential at the tree level.
Meanwhile, at the loop level, the Higgs potential is generated
by the AB effect due to the non-simplicity of spacetime.
This symmetry breaking mechanism is called the Hosotani
mechanism \cite{Hosotani:1983xw,Hosotani:1988bm}.

It has been found that the Higgs potential is finite, i.e.
independent of any UV cutoff at the one- and two-loop levels
in some GHU models
\cite{Hosotani:1983xw,Hosotani:1988bm,Davies:1988wt,
Antoniadis:2001cv,Maru:2006wa,Hosotani:2007kn,
Hisano:2019cxm,Nishikawa:2020fgg},
although higher-dimensional
gauge theories are generally non-renormalizable.
Moreover, the finiteness at all orders has been conjectured
\cite{vonGersdorff:2002rg,Hosotani:2005fk,Hosotani:2006nq}.

In the previous work \cite{Hisano:2019cxm},
we confirmed that the Higgs potential does not suffer from
the divergence at the two-loop level in a non-Abelian gauge
theory defined on $\mathbf M^4\times S^1$ spacetime,
where $\mathbf M^n$ is the $n$-dimensional Minkowski
spacetime with $n\geq1$ and $S^1$ is a circle.
Besides, we proceeded to discuss the finiteness
of the Higgs potential on $\mathbf M^5\times S^1$ spacetime,
which is related to this paper.
Evaluating the four-Fermi diagrams at the one-loop level,
we obtained the logarithmic divergences contributing to the Higgs
potential.
Hence, the Higgs potential is UV sensitive
on the six-dimensional spacetime.
This is consistent with the non-renormalizability of
higher-dimensional gauge theory.
Based on this result,
we concluded that the Higgs potential would be also suffering
from the divergence in the five-dimensional spacetime.

In this paper, we go back to $\mathbf M^4\times S^1$
spacetime again
and explicitly show that the Higgs potential depends on
UV theories in an $SU(\mathcal N)$ GHU model.
Since there are no logarithmic divergences at the one-loop level
in odd-dimensional theories,
we consider divergences at the two-loop level
in the dimensional regularization scheme.
The four-Fermi diagrams, in practice,
are evaluated at the two-loop level.
We find that they are indeed logarithmically divergent and their
counter terms contribute to the Higgs potential
at the four-loop level.
This fact means the UV sensitivity of the Higgs potential
in this GHU model.
Since we use a simple setup,
the Higgs potential would generically be UV sensitive
in other GHU models.

The remainder of this paper is organized as followes.
In Sect.~\ref{sec:hosotani_mec},
we briefly describe the Hosotani mechanism along with
the theoretical setup.
In Sect.~\ref{sec:main},
we explain how to evaluate the divergence of
the Higgs potential and show that the Higgs potential
receives a contribution from counter terms
to the divergences that cannot be subtracted
by the renormalization to the gauge coupling.
Finally,
we summarize the results obtained in this paper
in Sect.~\ref{sec:summary}.

\section{Hosotani mechanism}
\label{sec:hosotani_mec}
In this section,
we describe a theoretical setup
used in this paper
and review the Hosotani mechanism.

Since the Hosotani mechanism is a quantum effect
related to the global structure of spacetime,
let us consider $\mathbf M^4\times S^1$
as an example of non-simply connected spacetime,
where $\mathbf M^4$ and $S^1$ are
the four-dimensional Minkowski spacetime
and a circle with radius $R$ respectively.
We use coordinates $x^\mu$ with $\mu\in\{0,1,2,3\}$
for $\mathbf M^4$
and $y\in[0,2\pi R)$ for $S^1$.
As mentioned above,
on a spacetime with a hole,
the fifth component of the gauge boson, $A_5^a$, has
its VEV expressed by
\begin{equation}
    \langle A_5^a\rangle
    =\frac{\theta^a}{2\pi Rg},
\end{equation}
where $\theta^a$'s are the Yang-Mills AB phases around $S^1$
and $g$ is the coupling constant.
Here, $a$ denotes the group index.

Through this paper,
we use the background field method \cite{Abbott:1980hw}
for calculating the contributions
to the Higgs effective potential
and $A_5^a$'s are shifted by its VEV;
\begin{equation}
    A_5^a\to
    A_5^a+\frac{\theta^a}{2\pi Rg}.
\end{equation}

Due to compactified extra-dimension,
the boundary conditions on field functions are introduced.
We consider a massless Dirac fermion, $\psi$,
as only one type of matter field and
suppose $A_M^a$ and $\psi$ satisfy
\begin{align}
    A_M^a(x^\mu,y+2\pi R)&=A_M^a(x^\mu,y),
    \label{eq:bound1}\\
    \psi(x^\mu,y+2\pi R)&=e^{i\beta}\psi(x^\mu,y),
    \label{eq:bound2}
\end{align}
where $M\in\{0,1,2,3,5\}$ and $\beta\in[0,2\pi)$.

The Lagrangian we consider
with an $SU(\mathcal N)$ gauge symmetry is
\begin{equation}
    \mathcal L =
    -\frac{1}{4} F_{MN}^aF^{aMN}
    +\bar\psi i\gamma^MD_M\psi
    +\mathcal L_{\rm GF}+\mathcal L_{\rm ghost},
\end{equation}
where the gauge fixing terms, $\mathcal L_{\rm GF}$,
and the Faddeev-Popov ghost terms, $\mathcal L_{\rm ghost}$,
are given by
\begin{equation}
    \mathcal L_{\rm GF}=-\frac{1}{2}\mathcal F^a\mathcal F^a,\qquad
    \mathcal F^a\equiv\partial^MA_M^a+\frac{f^{abc}}{2\pi R}A^b_5\theta^c,
\end{equation}
\begin{equation}
    \mathcal L_{\rm ghost}=-\bar c^a
    \left[
    \partial^MD_M^{ab}
    -\frac{f^{ace}f^{bed}}{2\pi R}\theta^c\left(\frac{\theta^d}{2\pi R}+gA_5^d\right)
    \right]c^b.
\end{equation}
Here, $f^{abc}$ denotes
the structure constant of an $SU(\mathcal N)$ and
the covariant derivative for each field is defined by
\begin{align}
    D_Mc&\equiv\left(\partial_M-igA_M^aT^a-i\frac{\theta^aT^a}{2\pi R}\delta_M^5\right)c,\\
    D_M\psi&\equiv
    \left(
        \partial_M-igA^a_M\tau^a
        -i\frac{\theta^a\tau^a}{2\pi R}\delta_M^5
    \right)\psi,
\end{align}
where $[T^a]_{bc}=-if^{abc}$ and
$\tau^a$'s are representation matricies of $\psi$.

Let us see the Hosotani mechanism with the setup above.
In our model,
the gauge boson propagator is given by
\begin{equation}
    S_A(p^\mu,n)
    =\frac{-i\eta^{MN}}{p^\mu p_\mu-\left(\frac{n}{R}+\frac{\theta^aT^a}{2\pi R}\right)^2},
\end{equation}
where $n\in\mathbf Z$ denotes the Kaluza-Klein (KK) modes.
To shift the fifth component of a momentum,
we consider the following gauge transformations;
\begin{align}
    A_5(x^\mu,y)&\to e^{-i\frac{\theta^aT^a}{2\pi R}y} A_5(x^\mu,y)e^{i\frac{\theta^aT^a}{2\pi R}y}
    -\frac{\theta^aT^a}{2\pi Rg},
    \label{eq_gaway1}\\
    \psi_\ell(x^\mu,y)&\to e^{-i\frac{\theta^a\tau_\ell^a}{2\pi R}y}\psi_\ell(x^\mu,y),
    \label{eq_gaway2}
\end{align}
where $A_M=A_M^aT^a$.
Without any boundary conditions, arbitrary $\theta^a$ can be gauged away
by the above gauge transformations.
With Eqs.~\eqref{eq:bound1} and \eqref{eq:bound2}, in contrast,
the gauge transformations are restricted to be periodic on $S^1$
in order to keep the boundary conditions invariant.
Namely, removable $\theta^a$'s satisfy
\begin{equation}
    e^{i\theta^aT^a}=\mathbb I,
\end{equation}
where $\mathbb I$ is the identity matrix.
Remaining $\theta^a$'s become physical degrees of freedom.
Due to the gauge symmetry,
$\theta^a$ has no potential at the tree level.
Under the boundary conditions
which characterize the non-simplicity of spacetime, however,
its potential is generated by loop corrections
as shown in \cite{Hosotani:1983xw,Hosotani:1988bm,Davies:1988wt,
Antoniadis:2001cv,Maru:2006wa,Hosotani:2007kn,
Hisano:2019cxm,Nishikawa:2020fgg}.
When the minimum point of the potential is nonzero,
the gauge symmetry is dynamically broken and
the gauge bosons become massive.
This is how the Hosotani mechanism works.

\section{Higgs potential divergence at the four-loop level}
\label{sec:main}
Up to the two-loop level, we have found no divergence of
the Higgs potential in the previous works
\cite{Hosotani:1983xw,Hosotani:1988bm,Davies:1988wt,
Antoniadis:2001cv,Maru:2006wa,Hosotani:2007kn,
Hisano:2019cxm,Nishikawa:2020fgg}.
However, since the higher-dimensional gauge theory is
non-renormalizable, it is natural that the Higgs potential
receives $\theta$-dependent contributions from the infinite
number of counter terms.
Indeed, we have proven that the four-Fermi diagrams
contribute to the Higgs potential on $\mathbf M^5\times S^1$.
On the five-dimensional spacetime, it is expected that
they are logarithmically divergent at the two-loop level
and their counter terms contribute to the Higgs potential.
In this section, we show those divergences and
the UV dependence of the Higgs potential.
For automating calculations, we have used
a Mathematica package, \textit{FeynCalc}
\cite{Shtabovenko:2020gxv,Shtabovenko:2016sxi,Mertig:1990an}.

Let us consider a diagram shown below;
\begin{equation*}
    \fdiagram{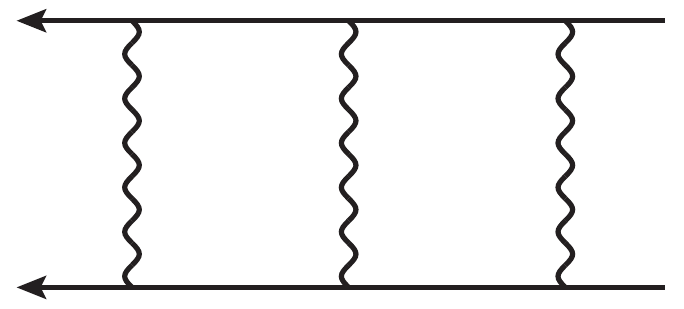},
\end{equation*}
which is logarithmically divergent.
To obtain its divergent part, we concentrate on loop momenta
going around the UV region.
Ignoring momenta lying external lines, we have
\begin{align}
    \left.\fdiagram{fig/4F2L-fig-01.pdf}\right|_{\rm div}
    &=\frac{1}{2\pi R}\sum_{n_1}\int\frac{d^4p}{(2\pi)^4}
    \frac{1}{2\pi R}\sum_{n_2}\int\frac{d^4k}{(2\pi)^4}\notag\\
    &\quad
    \times\left[ig\gamma^M\frac{i}{-\left(\gamma^\mu p_\mu-\gamma_5\frac{n_1}{R}\right)}ig\gamma^N\frac{i}{\gamma^\mu k_\mu-\gamma_5\frac{n_2}{R}}ig\gamma^L\right]_{\alpha\beta}[\tau^a\tau^b\tau^c]_{ij}\notag\\
    &\quad
    \times\frac{-i}{p^\mu p_\mu-\left(\frac{n_1}{R}\right)^2}
    \frac{-i}{(p+k)^\mu (p+k)_\mu-\left(\frac{n_1+n_2}{R}\right)^2}
    \frac{-i}{k^\mu k_\mu-\left(\frac{n_2}{R}\right)^2}\notag\\
    &\quad
    \times\left[ig\gamma_M\frac{i}{\gamma^\mu p_\mu-\gamma_5\frac{n_1}{R}}ig\gamma_N\frac{i}{-\left(\gamma^\mu k_\mu-\gamma_5\frac{n_2}{R}\right)}ig\gamma_L\right]_{\gamma\delta}[\tau^a\tau^b\tau^c]_{kl},
\end{align}
where $\alpha$ and $\gamma$ are spin indices of
$\bar\psi$, and $\beta$ and $\delta$ are those of $\psi$.
$i$, $j$, $k$, $l$ represent indices of $\tau^a$'s.
Note that all summation indices in this paper run all integers.
In the previous work~\cite{Hisano:2019cxm},
we derived the following formula%
\footnote{
    This transformation has been introduced in
    \cite{DaRold:2003yi}
    and used in
    \cite{Heffner:2015zna,Reinhardt:2016xci,Anber:2014sda,
    Ishikawa:2019tnw}
    for Abelian theory.
    For non-Abelian cases,
    the previous calculations
    \cite{Hosotani:1983xw,Hosotani:1988bm,Davies:1988wt}
    have adopted essentially the same way, while
    we use it before the four-dimensional integration
    to obtain the five-dimensional one.
};
\begin{equation}
    \frac{1}{2\pi R}\sum_n
    S\left(\frac{n}{R}+\frac{\Theta}{2\pi R}\right)
    =\sum_me^{i\Theta m}
    \int_{-\infty}^{\infty} \frac{dk_5}{2\pi}
    e^{-i2\pi Rk_5m}S(k_5),
    \label{eq:poisson}
\end{equation}
where $\Theta$ is an arbitrary Hermitian matrix.
Here, $S(\cdot)$ denotes an analytic function
and its generalization to a matrix-valued one.
Using Eq.~\eqref{eq:poisson} for $\Theta=0$,
we get
\begin{align}
    \left.\fdiagram{fig/4F2L-fig-01.pdf}\right|_{\rm div}
    &=-ig^6\sum_{m_1,m_2}\int\frac{d^5p}{(2\pi)^5}
    \int\frac{d^5k}{(2\pi)^5}e^{-i2\pi R(m_1p_5+m_2k_5)}\notag\\
    &\quad
    \times\frac{p^Pk^Qp^Rk^S}{(p^2)^3(p+k)^2(k^2)^3}\notag\\
    &\quad
    \times[\gamma^M\gamma_P\gamma^N\gamma_Q\gamma^L]_{\alpha\beta}
    [\gamma_M\gamma_R\gamma_N\gamma_S\gamma_L]_{\gamma\delta}\notag\\
    &\quad
    \times[\tau^a\tau^b\tau^c]_{ij}[\tau^a\tau^b\tau^c]_{kl}.
    \label{eq:4fermi}
\end{align}

Let us define $\mathcal I$ by
\begin{equation}
    \mathcal I 
    =\mathcal I(x,y)
    \equiv -\int\frac{d^5p}{(2\pi)^5}\int\frac{d^5k}{(2\pi)^5}
    \frac{1}{(p^2)^3(p+k)^2(k^2)^3}
    e^{-i2(p\cdot x+k\cdot y)},
\end{equation}
where $x$ and $y$ are space-like vectors.
For the spacetime dimension $D=5-2\epsilon$ with $\epsilon>0$,
$\mathcal I$ is calculated using a formula deduced in Appendix~\ref{sec:loop}.
Integrals in Eq.~\eqref{eq:4fermi} can be rewritten as a derivative of $\mathcal I$;
\begin{equation}
    \int\frac{d^5p}{(2\pi)^5}
    \int\frac{d^5k}{(2\pi)^5}
    \frac{p^Pk^Qp^Rk^S}{(p^2)^3(p+k)^2(k^2)^3}
    e^{-i2\pi R(m_1p_5+m_2k_5)}
    =-2^{-4}
    \partial_x^P\partial_y^Q\partial_x^R\partial_y^S
    \mathcal I\Bigr|_{\substack{x^M=\delta_5^Mm_1\pi R\\y^M=\delta_5^Mm_2\pi R}}.
    \label{eq:derivative}
\end{equation}
The behavior of the integrand in the UV region is shown in Eq.~\eqref{eq:int_p};
after integration over $k$ and angular variables,
the remaining (radial) integral has the form,
\begin{equation}
    \mathcal I
    \propto\int_0^\infty d|p_E|\,|p_E|^{a-1}K_r(2b\sqrt{-y^2}|p_E|)\,{}_0F_1\left(\frac{a+r}{2};(x-\beta y)^2|p_E|^2\right),
    \label{eq:pE_dep}
\end{equation}
where $a$, $b$, $r$ and $\beta$ are independent of $x$ and $y$.
Here, $K_r(z)$ is the modified Bessel function of the second kind and
${}_0F_1(a;z)$ is a generalized hypergeometric function.
Note that ${}_0F_1(a;z)$ is expressed by the Bessel function of the first kind, $J_\alpha(z)$;
\begin{equation}
    J_\alpha(2z)
    =\frac{(z)^\alpha}{\Gamma(\alpha+1)}\,{}_0F_1(\alpha+1;-z^2).
    \label{eq:Bessel_J}
\end{equation}
Plugging Eq.~\eqref{eq:Bessel_J} into Eq.~\eqref{eq:pE_dep},
we see that the integrand is a multiplication of $J_\alpha$, $K_r$, and the power of $|p_E|$.
Therefore, the UV divergence of $\mathcal I$ is suppressed by $K_r$ for its exponential dumping
when $y^M=\delta_5^Mm_2\pi R\neq0$.
Because $\mathcal I(x,y)=\mathcal I(y,x)$,
there is also no UV divergence when $x^M=\delta_5^Mm_1\pi R\neq0$.

To evaluate the UV divergence of $\mathcal I$,
we set $m_1=m_2=0$.
Substituting Eq.~\eqref{eq:final} into Eq.~\eqref{eq:derivative},
we obtain
\begin{multline}
    \left.\int\frac{d^Dp}{(2\pi)^D}
    \int\frac{d^Dk}{(2\pi)^D}
    \frac{p^Pk^Qp^Rk^S}{(p^2)^3(p+k)^2(k^2)^3}\right|_{\rm div}\\
    =-\frac{1}{2^4}\frac{1}{(4\pi)^5\cdot16\epsilon}
    \int_0^1d\alpha\int_0^1d\beta\,\alpha^{2}(1-\alpha)^{-4}(1-\beta)^{2}
    \left[\frac{\alpha}{1-\alpha}+\beta(1-\beta)\right]^{-\frac{9}{2}}\\
    \left.\times\partial_x^P\partial_y^Q\partial_x^R\partial_y^S
    \left[-(x-\beta y)^2-y^2\left(\frac{\alpha}{1-\alpha}+\beta(1-\beta)\right)\right]^{2}
    \right|_{x=y=0}.
\end{multline}
The derivatives are given by
\begin{multline}
    \partial_x^P\partial_y^Q\partial_x^R\partial_y^S
    \left[-(x-\beta y)^2-y^2\left(\frac{\alpha}{1-\alpha}+\beta(1-\beta)\right)\right]^{2}\\
    =8\beta^2(\eta^{PQ}\eta^{RS}+\eta^{PS}\eta^{QR})
    +8\eta^{PR}\eta^{QS}\left(\frac{\alpha}{1-\alpha}+\beta\right).
\end{multline}
In Appendix~\ref{sec:loop},
it is shown that
\begin{multline}
    \int_0^1d\alpha\int_0^1d\beta\,\alpha^{s-1}(1-\alpha)^{-s-1}\beta^{t-1}(1-\beta)^{u-1}
    \left[\frac{\alpha}{1-\alpha}+\beta(1-\beta)\right]^v\\
    =B(s,-s-v)B(s+t+v,s+u+v)
\end{multline}
with
\begin{equation}
    B(x,y)
    \equiv\int_0^1dt\,t^{x-1}(1-t)^{y-1}
    =\frac{\Gamma(x)\Gamma(y)}{\Gamma(x+y)}.
\end{equation}
Using this formula, we have
\begin{multline}
    \left.\fdiagram{fig/4F2L-fig-01.pdf}\right|_{\rm div}
    =\frac{ig^6}{2^{14}\cdot105\pi^4\epsilon}
    (\eta^{PQ}\eta^{RS}+\eta^{PS}\eta^{QR}+22\eta^{PR}\eta^{QS})\\
    \times[\gamma^M\gamma_P\gamma^N\gamma_Q\gamma^L]_{\alpha\beta}
    [\gamma_M\gamma_R\gamma_N\gamma_S\gamma_L]_{\gamma\delta}
    [\tau^a\tau^b\tau^c]_{ij}[\tau^a\tau^b\tau^c]_{kl}.
\end{multline}

Repeating the above procedure for the other two-loop four-Fermi diagrams,
we find the $\epsilon$-poles at the two-loop level.
They are shown explicitly in Appendix~\ref{sec:results}.
The divergence has the form,
\begin{equation}
    -\frac{ig^6}{\epsilon}\sum_{\mathcal X}
    C_{\mathcal X}
    [G_{\mathcal X}^{(1)}]_{\alpha\beta}
    [G_{\mathcal X}^{(2)}]_{\gamma\delta}
    [T_{\mathcal X}^{(1)}]_{ij}
    [T_{\mathcal X}^{(2)}]_{kl}
    -(\alpha\leftrightarrow\gamma,\, i\leftrightarrow k),
\end{equation}
where $\mathcal X$ denotes diagrams with four fermion legs
and $C_{\mathcal X}$ is a constant.
Here, $G_{\mathcal X}^{(1,2)}$ and $T_{\mathcal X}^{(1,2)}$
are products of $\gamma^M$'s and $\tau^a$'s respectively.
The above example corresponds to
\begin{equation}
    \mathcal X
    =\fdiagram{fig/4F2L-fig-01.pdf},
\end{equation}
\begin{align}
    G_{\mathcal X}^{(1)}
    &=\gamma^M\gamma_P\gamma^N\gamma_Q\gamma^L,\\
    G_{\mathcal X}^{(2)}
    &=\gamma_M\gamma_R\gamma_N\gamma_S\gamma_L,\\
    T_{\mathcal X}^{(1)}
    &=T_{\mathcal X}^{(2)}
    =\tau^a\tau^b\tau^c.
\end{align}

\begin{figure}
\centering
\includegraphics[width=14cm]{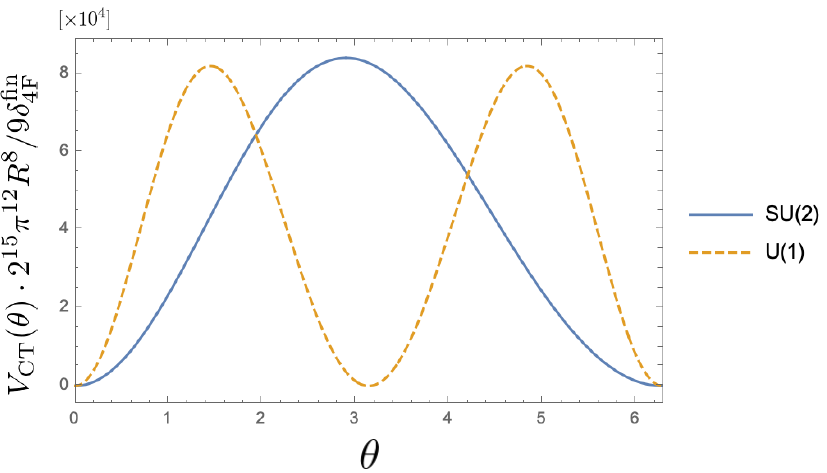}
\caption{
    Solid line:
    $V_{\rm CT}(\theta)$ in an $SU(2)$ gauge theory
    with a fermion in the fundamental representation.
    By $SU(2)$ transformations,
    the VEV of the Higgs boson can be written as a diagonal matrix;
    $\langle A_5\rangle={\rm diag}(\theta,-\theta)/2\pi Rg$.
    We have set $\beta$ to be zero.
    The contribution from the counter terms
    with the $\theta$-dependence results in
    the UV sensitivity of the Higgs potential.
    Dashed line:
    $V_{\rm CT}(\theta)$ in an Abelian gauge theory
    with a fermion whose $U(1)$-charge equals to one.
    $\beta$ is specified to be zero.
    In the previous work \cite{Hisano:2019cxm},
    it was shown that contributions to the Higgs potential from
    the one-loop four-Fermi diagrams vanished in the Abelian gauge theory.
    Based on this numerical calculation,
    we reject the all-order finiteness of the Higgs potential
    in an Abelian gauge theory.
}
\label{fig:Vct}
\end{figure}

The following counter term is introduced to cancel the above
divergence%
\footnote{
    We do not confirm $\mathcal L_{\rm CT}$ nonzero by computing it directly.
    When $\mathcal L_{\rm CT}=0$, it does not contribute to the Higgs potential.
    As a proof of the existence of $\mathcal L_{\rm CT}$ in a specific case,
    FIG.~\ref{fig:Vct} shows numerical results of its contributions to the Higgs potential.
};
\begin{equation}
    \mathcal L_{\rm CT}
    =\frac{\delta_{\rm 4F}}{2}
    \sum_{\mathcal X}C_{\mathcal X}
    (\bar\psi G_{\mathcal X}^{(1)}T_{\mathcal X}^{(1)}\psi)
    (\bar\psi G_{\mathcal X}^{(2)}T_{\mathcal X}^{(2)}\psi),
\end{equation}
where
$\delta_{\rm 4F}=\delta_{\rm 4F}^{\rm div}
+\delta_{\rm 4F}^{\rm fin}$.
Here, $\delta_{\rm 4F}^{\rm fin}$ is an arbitrary constant
and $\delta_{\rm 4F}^{\rm div}$ is defined as
\begin{equation}
    \delta_{\rm 4F}^{\rm div}
    =\frac{g^6}{\epsilon}
\end{equation}
to subtract the $\epsilon$-pole.

Closing the fermion lines, we get a contribution to
the Higgs potential from $\mathcal L_{\rm CT}$
with nontrivial $\theta$-dependence;
\begin{align}
    V_{\rm CT}(\theta)
    &=\frac{i}{2}
    \frac{1}{2\pi R}\sum_{n_1}\int\frac{d^4p}{(2\pi)^4}
    \frac{1}{2\pi R}\sum_{n_2}\int\frac{d^4k}{(2\pi)^4}\,
    i\delta_{\rm 4F}^{\rm fin}
    \sum_{\mathcal X}C_{\mathcal X}\notag\\
    &
    \times
    \left\{\tr\left[G_{\mathcal X}^{(1)}T_{\mathcal X}^{(1)}
    \frac{i}{\gamma_\mu p^\mu-\gamma_5\left(\frac{n_1}{R}+\frac{\theta^a\tau^a-\beta}{2\pi R}\right)}\right]
    \tr\left[G_{\mathcal X}^{(2)}T_{\mathcal X}^{(2)}
    \frac{i}{\gamma_\mu k^\mu-\gamma_5\left(\frac{n_2}{R}+\frac{\theta^a\tau^a-\beta}{2\pi R}\right)}\right]\right.\notag\\
    &\hspace{8ex}
    -\left.\tr\left[G_{\mathcal X}^{(1)}T_{\mathcal X}^{(1)}
    \frac{i}{\gamma_\mu p^\mu-\gamma_5\left(\frac{n_1}{R}+\frac{\theta^a\tau^a-\beta}{2\pi R}\right)}
    G_{\mathcal X}^{(2)}T_{\mathcal X}^{(2)}
    \frac{i}{\gamma_\mu k^\mu-\gamma_5\left(\frac{n_2}{R}+\frac{\theta^a\tau^a-\beta}{2\pi R}\right)}\right]\right\},
\end{align}
where we have traced out matrix indices of both $\gamma^M$'s
and $\tau^a$'s.
Using Eq.~\eqref{eq:poisson}, we get
\begin{align}
    V_{\rm CT}(\theta)
    &=\frac{\delta_{\rm 4F}^{\rm fin}}{2}
    \sum_{m_1,m_2}
    \int\frac{d^5p}{(2\pi)^5}\frac{p^A}{p^2}
    \int\frac{d^5k}{(2\pi)^5}\frac{k^B}{k^2}
    e^{-i2\pi R(m_1p_5+m_2k_5)}
    \sum_{\mathcal X}C_{\mathcal X}\notag\\
    &
    \times
    \left\{\tr\left[G_{\mathcal X}^{(1)}T_{\mathcal X}^{(1)}
    \gamma_Ae^{i(\theta^a\tau^a-\beta)m_1}\right]
    \tr\left[G_{\mathcal X}^{(2)}T_{\mathcal X}^{(2)}
    \gamma_Be^{i(\theta^a\tau^a-\beta)m_2}\right]\right.\notag\\
    &\hspace{8ex}
    -\left.\tr\left[G_{\mathcal X}^{(1)}T_{\mathcal X}^{(1)}
    \gamma_Ae^{i(\theta^a\tau^a-\beta)m_1}
    G_{\mathcal X}^{(2)}T_{\mathcal X}^{(2)}
    \gamma_Be^{i(\theta^a\tau^a-\beta)m_2}\right]\right\}.
\end{align}
In the previous work \cite{Hisano:2019cxm},
we have derived that
\begin{equation}
    \int\frac{d^Dk}{(2\pi)^D}(-k^2-i\epsilon)^{-s}e^{-i2k\cdot x}
    =\frac{i}{(4\pi)^{D/2}}\frac{\Gamma(\frac{D}{2}-s)}{\Gamma(s)}(-x^Mx_M)^{s-D/2}.
\end{equation}
From this formula, we obtain
\begin{align}
    V_{\rm CT}(\theta)
    &=\frac{9\delta_{\rm 4F}^{\rm fin}}{2^{15}\pi^{12}R^8}
    \sum_{m_1\neq0}\sum_{m_2\neq0}
    \frac{m_1m_2}{|m_1|^5|m_2|^5}
    \sum_{\mathcal X}C_{\mathcal X}\notag\\
    &
    \times
    \left\{\tr\left[G_{\mathcal X}^{(1)}T_{\mathcal X}^{(1)}
    \gamma_5e^{i(\theta^a\tau^a-\beta)m_1}\right]
    \tr\left[G_{\mathcal X}^{(2)}T_{\mathcal X}^{(2)}
    \gamma_5e^{i(\theta^a\tau^a-\beta)m_2}\right]\right.\notag\\
    &\hspace{8ex}
    -\left.\tr\left[G_{\mathcal X}^{(1)}T_{\mathcal X}^{(1)}
    \gamma_5e^{i(\theta^a\tau^a-\beta)m_1}
    G_{\mathcal X}^{(2)}T_{\mathcal X}^{(2)}
    \gamma_5e^{i(\theta^a\tau^a-\beta)m_2}\right]\right\}.
\end{align}
The contributions from each diagram are explicitly written
down in Appendix~\ref{sec:results}.
To get $V_{\rm CT}(\theta)$ in an Abelian gauge theory,
we replace $\tau^a$'s with $Q$, the $U(1)$-charge of $\psi$.
We have computed $V_{\rm CT}(\theta)$ in
an $SU(2)$ gauge theory with a fermion in the fundamental representation
and an Abelian case with a fermion having the $U(1)$-charge $Q=1$,
which is shown in FIG.~\ref{fig:Vct}.
Therefore, it is concluded that the $\theta$-dependent part
of the Higgs potential is UV sensitive.

\section{Summary}
\label{sec:summary}
In this paper, we investigate the finiteness of the Higgs
potential beyond the two-loop level in the GHU by evaluating
the loop corrections explicitly.
While the Higgs potential was found finite at the one- or two-loop levels
on many non-simply connected manifolds, its finiteness at
higher-order had been unclear.
As suggested in the previous work \cite{Hisano:2019cxm},
it is shown that
the Higgs potential receives the nontrivial $\theta$-dependent
contributions from the counter terms for the four-Fermi
diagrams on $\mathbf M^4\times S^1$ at the four-loop level
and, thus, it is UV sensitive.

For logarithmic divergences found in this paper,
when we impose a UV cutoff on the GHU to make sense as
an effective field theory, the maximum value of the cutoff,
denoted as $\Lambda_{\rm max}$, satisfies
$\ln(R\Lambda_{\rm max})\sim g^2\Lambda_{\rm max}$.
Hence, in the GHU, the perturbation is valid at most around
the compactification scale.

\begin{acknowledgments}
The author A.Y. thanks Junji Hisano for
his tremendous supports and meaningful discussions
and also thanks Yutaro Shoji for his advice and
unconditional dedication to improving this paper.
\end{acknowledgments}

\appendix

\section{Two-loop integrals}
\label{sec:loop}
This appendix is dedicated to evaluating a following integral;
\begin{equation}
    \mathcal I 
    \equiv \int\frac{d^Dp}{(2\pi)^D}\int\frac{d^Dk}{(2\pi)^D}
    [-p^2-i\epsilon]^{-s}[-(p+k)^2-i\epsilon]^{-t}[-k^2-i\epsilon]^{-u}e^{-i2(p\cdot x+k\cdot y)},
\end{equation}
where $s$, $t$, $u$ are positive constants satisfying $s+t+u>D$
and $x$, $y$ are space-like vectors independent of $p$ and $k$.

Introducing the Feynman parameters,
we get
\begin{multline}
\mathcal I
=\int\frac{d^Dp}{(2\pi)^D}\int\frac{d^Dk}{(2\pi)^D}
\int_0^1d\alpha\int_0^{1-\alpha}d\beta\\
\times\frac{\alpha^{s-1}\beta^{t-1}(1-\alpha-\beta)^{u-1}}{[-\alpha p^2-\beta(p+k)^2-(1-\alpha-\beta)k^2-i\epsilon]^{s+t+u}}
\frac{\Gamma(s+t+u)}{\Gamma(s)\Gamma(t)\Gamma(u)}
e^{-i2(p\cdot x+k\cdot y)}.
\end{multline}

In the previous work \cite{Hisano:2019cxm}, we showed
\begin{multline}
\int\frac{d^Dk}{(2\pi)^D}\,(-k^2+2p\cdot k+m^2-i\epsilon)^{-s}e^{-i2k\cdot x}\\
=\frac{2i}{(4\pi)^{D/2}\Gamma(s)}
\frac{e^{-i2p\cdot x}(-x^2)^{s/2-D/4}}{(p^2+m^2-i\epsilon)^{s/2-D/4}}
K_{s-D/2}(2\sqrt{(-x^2)(p^2+m^2-i\epsilon)}),
\end{multline}
where $\mathrm{Re}(s)>0$, $p^2+m^2\neq0$.
Here, $K_r(z)$ is the modified Bessel function of the second kind.
Using this formula, we have
\begin{align}
\mathcal I
&=\frac{2i(-y^2)^{\frac{s+t+u}{2}-\frac{D}{4}}}{(4\pi)^{D/2}\Gamma(s)\Gamma(t)\Gamma(u)}
\int_0^1d\alpha\int_0^1d\beta\int\frac{d^Dp}{(2\pi)^D}\notag\\
&\hspace{10ex}
\times\alpha^{s-1}(1-\alpha)^{-s-1}\beta^{t-1}(1-\beta)^{u-1}
e^{-i2p\cdot(x-\beta y)}\notag\\
&\hspace{10ex}\times
\left\{-\left[\frac{\alpha}{1-\alpha}+\beta(1-\beta)\right]p^2-\frac{i\epsilon}{1-\alpha}\right\}^{-\frac{s+t+u}{2}+\frac{D}{4}}
\notag\\
&\hspace{10ex}\times K_{s+t+u-\frac{D}{2}}
\left(2\sqrt{(-y^2)\left\{-\left[\frac{\alpha}{1-\alpha}+\beta(1-\beta)\right]p^2-\frac{i\epsilon}{1-\alpha}\right\}}\right),
\end{align}
where we have scaled $\beta$ to $(1-\alpha)\beta$.

There exist at most two mutually different solutions $\beta\in(0,1)$ for $(x-\beta y)^2=0$.
When the number of such solutions is denoted by $n$,
we divide the integration interval over $\beta$ into $n+1$ parts
where $(x-\beta y)^2$ is positive or negative entirely;
\begin{equation}
    \int_0^1d\beta\,\beta^{t-1}(1-\beta)^{u-1}\mathcal F
    =\sum_{i=0}^n\int_{\beta_{i}}^{\beta_{i+1}}d\beta\,\beta^{t-1}(1-\beta)^{u-1}\mathcal F,
\end{equation}
where $\beta_0\equiv0$, $\beta_{n+1}\equiv1$, and
$\beta_i\in(0,1)$ satisfies $(x-\beta_iy)^2=0$ for $0<i<n+1$.
Here, we have defined
\begin{multline}
    \mathcal F
    \equiv\int\frac{d^Dp}{(2\pi)^D}\,e^{-i2p\cdot(x-\beta y)}
    \left\{-\left[\frac{\alpha}{1-\alpha}+\beta(1-\beta)\right]p^2-\frac{i\epsilon}{1-\alpha}\right\}^{-\frac{s+t+u}{2}+\frac{D}{4}}\\
    \times K_{s+t+u-\frac{D}{2}}\left(2\sqrt{(-y^2)\left\{-\left[\frac{\alpha}{1-\alpha}
    +\beta(1-\beta)\right]p^2-\frac{i\epsilon}{1-\alpha}\right\}}\right).
\end{multline}

When $x-\beta y$ is space-like,
we set $x^0-\beta y^0$ to be zero via the Lorentz transformations.
Using the Wick rotation,
\begin{equation}
p^0\to ip_E^0,\qquad
\mathbf p\equiv\mathbf p_E,
\end{equation}
\begin{equation}
\mathbf x\equiv\mathbf x_E,\qquad
\mathbf y\equiv\mathbf y_E,
\end{equation}
we get
\begin{multline}
    \mathcal F
    =\frac{i}{(2\pi)^D}\int_0^\infty d|p_E|\,|p_E|^{D-1}
    \left\{\left[\frac{\alpha}{1-\alpha}+\beta(1-\beta)\right]p_E^2\right\}^{-\frac{s+t+u}{2}+\frac{D}{4}}\\
    \times K_{s+t+u-\frac{D}{2}}\left(2|y_E|\sqrt{\left[\frac{\alpha}{1-\alpha}+\beta(1-\beta)\right]p_E^2}\right)
    \int d\Omega_D\,e^{i2\mathbf p_E\cdot(\mathbf x_E-\beta\mathbf y_E)}.
\end{multline}
Carrying out the integral over all angles except $\theta$, the angle between $p_E=(p_E^0,\mathbf p_E)$
and $x_E-\beta y_E=(0,\mathbf x_E-\beta\mathbf y_E)$, we get
\begin{equation}
\int d\Omega_D\,e^{i2\mathbf p_E\cdot(\mathbf x_E-\beta\mathbf y_E)}
=\Omega_{D-1}\int_0^\pi d\theta\, \sin^{D-2}\theta\,e^{i2|p_E||x_E-\beta y_E|\cos\theta},
\label{eq:int_angles}
\end{equation}
where $\Omega_D$ is the area of the unit sphere in the $D$-dimensional space;
\begin{equation}
\Omega_D
=\frac{2\pi^{D/2}}{\Gamma(D/2)}.
\end{equation}
The integral over $\theta$ is evaluated as
\begin{equation}
\int_0^\pi d\theta\, \sin^{a-1}\theta\,e^{i2b\cos\theta}
=\frac{\sqrt\pi\Gamma\left(\frac{a}{2}\right)}{\Gamma\left(\frac{a+1}{2}\right)}\,{}_0F_1\left(\frac{a+1}{2};-b^2\right)
\label{eq:int_theta}
\end{equation}
for $\mathrm{Re}(a)>0$ and $b\in \mathbf R$.
Here, ${}_0F_1(a;z)$ is a generalized hypergeometric function;
\begin{equation}
{}_0F_1(a;z)
\equiv \sum_{n=0}^\infty\frac{1}{(a)_n}\frac{z^n}{n!},
\end{equation}
where
\begin{equation}
(a)_0=1,\qquad
(a)_n=\prod_{m=0}^{n-1}(a+m),\qquad
n\in\mathbf Z^+.
\end{equation}

When $x-\beta y$ is time-like,
on the other hand,
$\mathbf{x}-\beta\mathbf{y}$ is set to be zero via the Lorentz transformations.
After the Wick rotation, $\mathcal F$ becomes
\begin{multline}
    \mathcal F
    =\frac{i}{(2\pi)^D}\int_0^\infty d|p_E|\,|p_E|^{D-1}
    \left\{\left[\frac{\alpha}{1-\alpha}+\beta(1-\beta)\right]p_E^2\right\}^{-\frac{s+t+u}{2}+\frac{D}{4}}\\
    \times K_{s+t+u-\frac{D}{2}}\left(2|y_E|\sqrt{\left[\frac{\alpha}{1-\alpha}+\beta(1-\beta)\right]p_E^2}\right)
    \int d\Omega_D\,e^{2p_E^0(x_E^0-\beta y_E^0)},
\end{multline}
where $x_E^0\equiv x^0$ and $y_E^0\equiv y^0$.
We evaluate angle integrals with a similar way to
Eqs.~\eqref{eq:int_angles} and \eqref{eq:int_theta};
\begin{equation}
    \int d\Omega_D\,e^{2p_E^0(x_E^0-\beta y_E^0)}
    =\Omega_{D-1}\int_0^\pi d\theta\, \sin^{D-2}\theta\,e^{2|p_E||x_E-\beta y_E|\cos\theta}
\end{equation}
and
\begin{equation}
    \int_0^\pi d\theta\, \sin^{a-1}\theta\,e^{2b\cos\theta}
    =\frac{\sqrt\pi\Gamma\left(\frac{a}{2}\right)}{\Gamma\left(\frac{a+1}{2}\right)}\,{}_0F_1\left(\frac{a+1}{2};b^2\right)
\end{equation}
for $\mathrm{Re}(a)>0$ and $b\in \mathbf R$.

The integral over $|p_E|$ is given by
\begin{equation}
\int_0^\infty d|p_E|\,|p_E|^{a-1}K_r(2b|p_E|)\,{}_0F_1\left(\frac{a+r}{2};c|p_E|^2\right)
=\frac{1}{4}\Gamma\left(\frac{a-r}{2}\right)\Gamma\left(\frac{a+r}{2}\right)b^{-r}(b^2-c)^{-\frac{a-r}{2}}
\label{eq:int_p}
\end{equation}
for $b>0$, $c<0$, $\mathrm{Re}(a-r)>0$, and $\mathrm{Re}(a+r)>0$.

Applying the above formulae to $\mathcal F$, we obtain
\begin{multline}
    \mathcal F
    =\frac{i}{2}\frac{\Gamma(D-s-t-u)}{(4\pi)^{D/2}}\left(\sqrt{-y^2}\right)^{\frac{D}{2}-s-t-u}
    \left[\frac{\alpha}{1-\alpha}+\beta(1-\beta)\right]^{\frac{D}{2}-s-t-u}\\
    \times\left[-(x-\beta y)^2-y^2\left(\frac{\alpha}{1-\alpha}+\beta(1-\beta)\right)\right]^{s+t+u-D}
\end{multline}
regardless of whether $x-\beta y$ is space-like or time-like.
Therefore, after integration over $k$ and $p$,
we get
\begin{multline}
\mathcal I
=-\frac{\Gamma(D-s-t-u)}{(4\pi)^D\Gamma(s)\Gamma(t)\Gamma(u)}
\int_0^1d\alpha\int_0^1d\beta\,\alpha^{s-1}(1-\alpha)^{-s-1}\beta^{t-1}(1-\beta)^{u-1}\\
\times\left[\frac{\alpha}{1-\alpha}+\beta(1-\beta)\right]^{\frac{D}{2}-s-t-u}
\left[-(x-\beta y)^2-y^2\left(\frac{\alpha}{1-\alpha}+\beta(1-\beta)\right)\right]^{s+t+u-D}.
\label{eq:final}
\end{multline}

By expanding the last factor,
finding the $\epsilon$-pole of $\mathcal I$ comes down to
calculating the following integral;
\begin{equation}
\mathcal J
\equiv\int_0^1d\alpha\int_0^1d\beta\,\alpha^{\sigma-1}(1-\alpha)^{-\sigma-1}\beta^{\tau-1}(1-\beta)^{\kappa-1}
\left[\frac{\alpha}{1-\alpha}+\beta(1-\beta)\right]^\lambda,
\end{equation}
where $\sigma$, $\tau$, $\kappa$, and $\lambda$ are arbitrary constants.

Setting $\gamma$ to be
\begin{equation}
\gamma
\equiv\frac{\alpha}{1-\alpha},
\end{equation}
we have
\begin{align}
\mathcal J
&=\int_0^\infty d\gamma\int_0^1d\beta\,\gamma^{\sigma-1}\beta^{\tau-1}(1-\beta)^{\kappa-1}
[\gamma+\beta(1-\beta)]^\lambda\notag\\
&=B(\sigma,-\sigma-\lambda)\int_0^1d\beta\,\beta^{\sigma+\tau+\lambda-1}(1-\beta)^{\sigma+\kappa+\lambda-1}\notag\\
&=B(\sigma,-\sigma-\lambda)B(\sigma+\tau+\lambda,\sigma+\kappa+\lambda),
\end{align}
where $B(x,y)$ is the beta function;
\begin{equation}
B(x,y)
\equiv\int_0^1dt\,t^{x-1}(1-t)^{y-1}
=\frac{\Gamma(x)\Gamma(y)}{\Gamma(x+y)}.
\end{equation}

\section{Table of divergence and its contribution to the Higgs potential}
\label{sec:results}
As shown in Sect.~\ref{sec:main},
at the two-loop level,
the divergences of the four-Fermi diagrams have the form,
\begin{equation}
    -\frac{ig^6}{\epsilon}
    C_{\mathcal X}
    [G_{\mathcal X}^{(1)}]_{\alpha\beta}
    [G_{\mathcal X}^{(2)}]_{\gamma\delta}
    [T_{\mathcal X}^{(1)}]_{ij}
    [T_{\mathcal X}^{(2)}]_{kl}
    -(\alpha\leftrightarrow\gamma,i\leftrightarrow k),
\end{equation}
where $G_{\mathcal X}^{(1,2)}$ and $T_{\mathcal X}^{(1,2)}$
are products of $\gamma^M$'s and $\tau^a$'s respectively.
Here, $C_{\mathcal X}$ is a constant and
$\mathcal X$ denotes divergent diagrams without
crossing of the fermion lines.
The second term represents the same diagram
with the fermion lines being crossed.

In this appendix, we summarize the divergent diagrams and
their contributions to the Higgs potential in a table.
For making appearance simple,
we write down the following factors;
\begin{equation}
    -64(4\pi)^5C_{\mathcal X}
    [G_{\mathcal X}^{(1)}]_{\alpha\beta}
    [G_{\mathcal X}^{(2)}]_{\gamma\delta}
    [T_{\mathcal X}^{(1)}]_{ij}
    [T_{\mathcal X}^{(2)}]_{kl}
\end{equation}
and
\begin{align}
    &-64(4\pi)^5C_{\mathcal X}
    \left\{\tr\left[G_{\mathcal X}^{(1)}T_{\mathcal X}^{(1)}
    \gamma_5e^{i(\theta^a\tau^a-\beta)m_1}\right]
    \tr\left[G_{\mathcal X}^{(2)}T_{\mathcal X}^{(2)}
    \gamma_5e^{i(\theta^a\tau^a-\beta)m_2}\right]\right.\notag\\
    &\hspace{8ex}
    -\left.\tr\left[G_{\mathcal X}^{(1)}T_{\mathcal X}^{(1)}
    \gamma_5e^{i(\theta^a\tau^a-\beta)m_1}
    G_{\mathcal X}^{(2)}T_{\mathcal X}^{(2)}
    \gamma_5e^{i(\theta^a\tau^a-\beta)m_2}\right]\right\}.
\end{align}
\newpage
\begin{longtable}[c]{cl}
    \caption{
        The divergent part of
        the four-Fermi diagrams at the two-loop level
        and its contribution to the Higgs potential.
        Each diagram is represented by $\mathcal X$.
        $G_{\mathcal X}^{(1)}$ and $T_{\mathcal X}^{(1)}$
        are products of $\gamma^M$'s and $\tau^a$'s,
        respectively, on the upper fermion line.
        $G_{\mathcal X}^{(2)}$ and $T_{\mathcal X}^{(2)}$
        are ones on the lower fermion line.
        $C_{\mathcal X}$ is a collection of
        the other factors
        up to $-ig^6/\epsilon$, where
        $\epsilon=(5-D)/2$ with the spacetime dimension $D$.
        Note that we have
        explicitly traced out with respect to $\gamma^M$'s
        in the expressions of the contributions
        to the Higgs potential.
        $c_r$ is a constant which depends on
        the representation of a fermion;
        $\tr[\tau^a\tau^b]=c_r\delta^{ab}$.
        For the fundamental representation, $c_r=1/2$.
    }
    \\\hline
    $\mathcal X$
    \quad&
    \begin{tabular}{l}
        $-64(4\pi)^5C_{\mathcal X}
        [G_{\mathcal X}^{(1)}]_{\alpha\beta}
        [G_{\mathcal X}^{(2)}]_{\gamma\delta}
        [T_{\mathcal X}^{(1)}]_{ij}
        [T_{\mathcal X}^{(2)}]_{kl}$,
        \\\\
        $-64(4\pi)^5C_{\mathcal X}
        \left\{\tr\left[G_{\mathcal X}^{(1)}T_{\mathcal X}^{(1)}
        \gamma_5e^{i(\theta^a\tau^a-\beta)m_1}\right]
        \tr\left[G_{\mathcal X}^{(2)}T_{\mathcal X}^{(2)}
        \gamma_5e^{i(\theta^a\tau^a-\beta)m_2}\right]\right.$
        \\\quad\hfill
        $-\left.\tr\left[G_{\mathcal X}^{(1)}T_{\mathcal X}^{(1)}
        \gamma_5e^{i(\theta^a\tau^a-\beta)m_1}
        G_{\mathcal X}^{(2)}T_{\mathcal X}^{(2)}
        \gamma_5e^{i(\theta^a\tau^a-\beta)m_2}\right]\right\}$
    \end{tabular}
    \\\hline\hline
    \endfirsthead
    \hline
    \endlastfoot
    \fdiagram{fig/4F2L-fig-01.pdf}
    \quad&
    \begin{tabular}{l}
        $\frac{4}{105} \pi \left(\eta^{PS} \eta^{QR}+22 \eta^{PR} \eta^{QS}+\eta^{PQ} \eta^{RS}\right)$
        \\\quad\hfill
        $\times\left[\gamma ^M\gamma _P\gamma ^N\gamma _Q\gamma ^L\right]{}_{\alpha \beta } \left[\gamma _M\gamma _R\gamma _N\gamma _S\gamma _L\right]{}_{\gamma \delta } \left[\tau^a\tau^b\tau^c\right]{}_{ij} \left[\tau^a\tau^b\tau^c\right]{}_{kl}$,
        \\\\
        $-\frac{128}{15} \pi \left\{394 \tr\left[\tau^a\tau^b\tau^ce^{i(\theta^d\tau^d-\beta)m_1}\right] \tr\left[\tau^a\tau^b\tau^ce^{i(\theta^d\tau^d-\beta)m_2}\right]\right.$
        \\\quad\hfill
        $\left.+393 \tr\left[\tau^a\tau^b\tau^ce^{i(\theta^d\tau^d-\beta)m_1}\tau^a\tau^b\tau^ce^{i(\theta^d\tau^d-\beta)m_2}\right]\right\}$
    \end{tabular}
    \\\hline
    \fdiagram{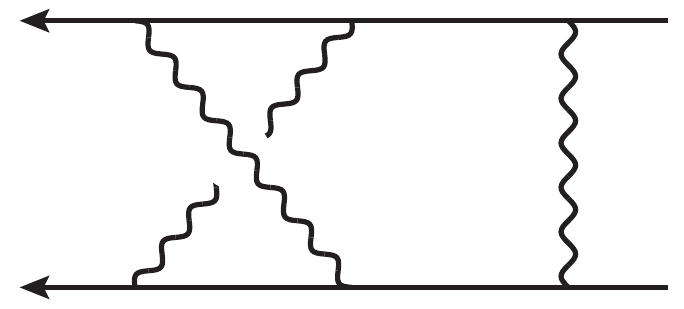}
    \quad&
    \begin{tabular}{l}
        $\frac{8}{105} \pi \left(\eta^{PS} \eta^{QR}-6 \eta^{PR} \eta^{QS}+\eta^{PQ} \eta^{RS}\right)$
        \\\quad\hfill
        $\times\left[\gamma ^M\gamma _P\gamma ^N\gamma _Q\gamma ^L\right]{}_{\alpha \beta } \left[\gamma _N\gamma _R\gamma _M\gamma _S\gamma _L\right]{}_{\gamma \delta } \left[\tau^a\tau^b\tau^c\right]{}_{ij} \left[\tau^b\tau^a\tau^c\right]{}_{kl}$,
        \\\\
        $\frac{128}{15} \pi \left\{80 \tr\left[\tau^a\tau^b\tau^ce^{i(\theta^d\tau^d-\beta)m_1}\right] \tr\left[\tau^b\tau^a\tau^ce^{i(\theta^d\tau^d-\beta)m_2}\right]\right.$
        \\\quad\hfill
        $\left.+81 \tr\left[\tau^a\tau^b\tau^ce^{i(\theta^d\tau^d-\beta)m_1}\tau^b\tau^a\tau^ce^{i(\theta^d\tau^d-\beta)m_2}\right]\right\}$
    \end{tabular}
    \\\hline
    \fdiagram{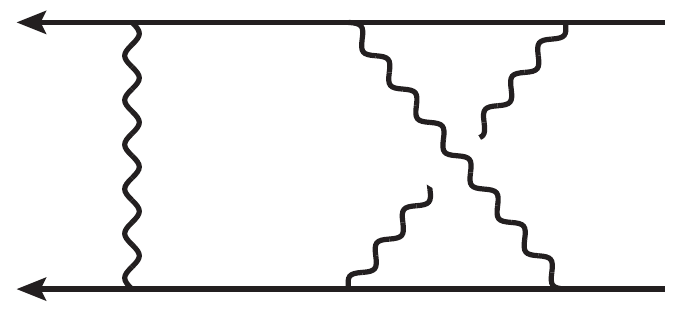}
    \quad&
    \begin{tabular}{l}
        $\frac{8}{105} \pi \left(\eta^{PS} \eta^{QR}-6 \eta^{PR} \eta^{QS}+\eta^{PQ} \eta^{RS}\right)$
        \\\quad\hfill
        $\times\left[\gamma ^M\gamma _P\gamma ^N\gamma _Q\gamma ^L\right]{}_{\alpha \beta } \left[\gamma _M\gamma _R\gamma _L\gamma _S\gamma _N\right]{}_{\gamma \delta } \left[\tau^a\tau^b\tau^c\right]{}_{ij} \left[\tau^a\tau^c\tau^b\right]{}_{kl}$,
        \\\\
        $\frac{128}{15} \pi \left\{80 \tr\left[\tau^a\tau^b\tau^ce^{i(\theta^d\tau^d-\beta)m_1}\right] \tr\left[\tau^a\tau^c\tau^be^{i(\theta^d\tau^d-\beta)m_2}\right]\right.$
        \\\quad\hfill
        $\left.+81 \tr\left[\tau^a\tau^b\tau^ce^{i(\theta^d\tau^d-\beta)m_1}\tau^a\tau^c\tau^be^{i(\theta^d\tau^d-\beta)m_2}\right]\right\}$
    \end{tabular}
    \\\hline
    \fdiagram{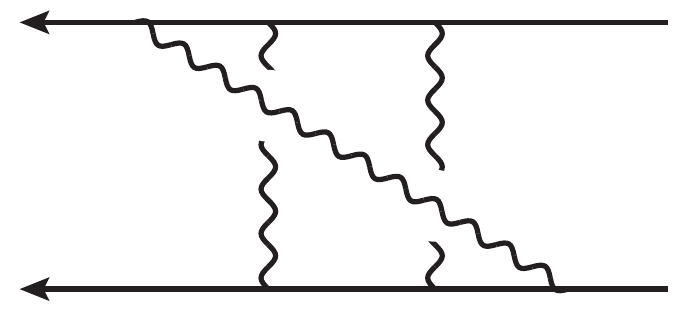}
    \quad&
    \begin{tabular}{l}
        $-\frac{8}{105} \pi \left(6 \eta^{PS} \eta^{QR}-\eta^{PR} \eta^{QS}-\eta^{PQ} \eta^{RS}\right)$
        \\\quad\hfill
        $\times\left[\gamma ^M\gamma _P\gamma ^N\gamma _Q\gamma ^L\right]{}_{\alpha \beta } \left[\gamma _N\gamma _R\gamma _L\gamma _S\gamma _M\right]{}_{\gamma \delta } \left[\tau^a\tau^b\tau^c\right]{}_{ij} \left[\tau^b\tau^c\tau^a\right]{}_{kl}$,
        \\\\
        $\frac{128}{15} \pi \left\{80 \tr\left[\tau^a\tau^b\tau^ce^{i(\theta^d\tau^d-\beta)m_1}\right] \tr\left[\tau^b\tau^c\tau^ae^{i(\theta^d\tau^d-\beta)m_2}\right]\right.$
        \\\quad\hfill
        $\left.+99 \tr\left[\tau^a\tau^b\tau^ce^{i(\theta^d\tau^d-\beta)m_1}\tau^b\tau^c\tau^ae^{i(\theta^d\tau^d-\beta)m_2}\right]\right\}$
    \end{tabular}
    \\\hline
    \fdiagram{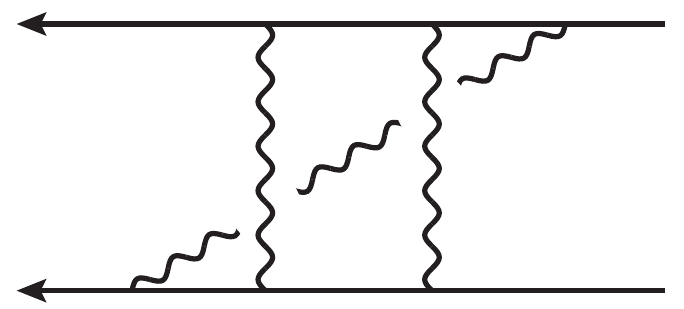}
    \quad&
    \begin{tabular}{l}
        $-\frac{8}{105} \pi \left(6 \eta^{PS} \eta^{QR}-\eta^{PR} \eta^{QS}-\eta^{PQ} \eta^{RS}\right)$
        \\\quad\hfill
        $\times\left[\gamma ^M\gamma _P\gamma ^N\gamma _Q\gamma ^L\right]{}_{\alpha \beta } \left[\gamma _L\gamma _R\gamma _M\gamma _S\gamma _N\right]{}_{\gamma \delta } \left[\tau^a\tau^b\tau^c\right]{}_{ij} \left[\tau^c\tau^a\tau^b\right]{}_{kl}$,
        \\\\
        $\frac{128}{15} \pi \left\{80 \tr\left[\tau^a\tau^b\tau^ce^{i(\theta^d\tau^d-\beta)m_1}\right] \tr\left[\tau^c\tau^a\tau^be^{i(\theta^d\tau^d-\beta)m_2}\right]\right.$
        \\\quad\hfill
        $\left.+99 \tr\left[\tau^a\tau^b\tau^ce^{i(\theta^d\tau^d-\beta)m_1}\tau^c\tau^a\tau^be^{i(\theta^d\tau^d-\beta)m_2}\right]\right\}$
    \end{tabular}
    \\\hline
    \fdiagram{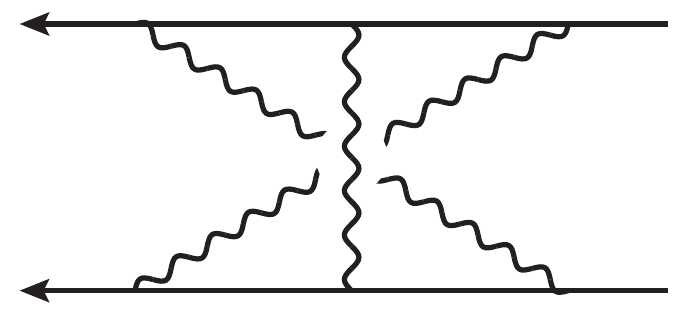}
    \quad&
    \begin{tabular}{l}
        $\frac{4}{105} \pi \left(22 \eta^{PS} \eta^{QR}+\eta^{PR} \eta^{QS}+\eta^{PQ} \eta^{RS}\right)$
        \\\quad\hfill
        $\times\left[\gamma ^M\gamma _P\gamma ^N\gamma _Q\gamma ^L\right]{}_{\alpha \beta } \left[\gamma _L\gamma _R\gamma _N\gamma _S\gamma _M\right]{}_{\gamma \delta } \left[\tau^a\tau^b\tau^c\right]{}_{ij} \left[\tau^c\tau^b\tau^a\right]{}_{kl}$,
        \\\\
        $-\frac{256}{15} \pi \left\{197 \tr\left[\tau^a\tau^b\tau^ce^{i(\theta^d\tau^d-\beta)m_1}\right] \tr\left[\tau^c\tau^b\tau^ae^{i(\theta^d\tau^d-\beta)m_2}\right]\right.$
        \\\quad\hfill
        $\left.+54 \tr\left[\tau^a\tau^b\tau^ce^{i(\theta^d\tau^d-\beta)m_1}\tau^c\tau^b\tau^ae^{i(\theta^d\tau^d-\beta)m_2}\right]\right\}$
    \end{tabular}
    \\\hline
    \fdiagram{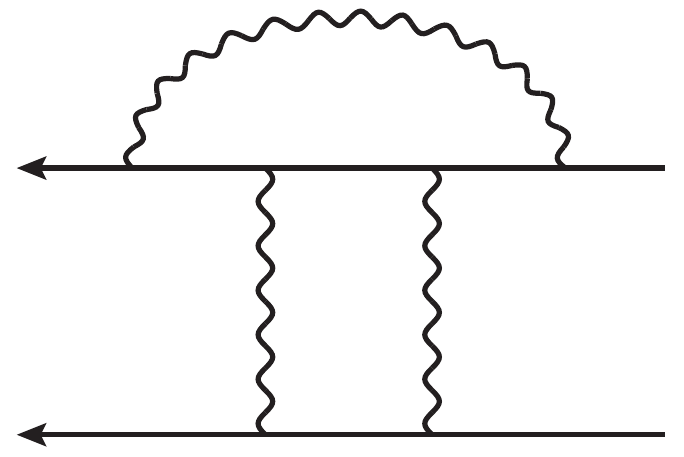}
    \quad&
    \begin{tabular}{l}
        $\frac{4}{35} \pi \left(\eta^{PS} \eta^{QR}-6 \eta^{PR} \eta^{QS}+\eta^{PQ} \eta^{RS}\right)$
        \\\quad\hfill
        $\times\left[\gamma ^M\gamma _P\gamma ^N\gamma _Q\gamma ^L\gamma _R\gamma _M\right]{}_{\alpha \beta } \left[\gamma _N\gamma _S\gamma _L\right]{}_{\gamma \delta } \left[\tau^a\tau^b\tau^c\tau^a\right]{}_{ij} \left[\tau^b\tau^c\right]{}_{kl}$,
        \\\\
        $\frac{1728}{5} \pi \left\{4 \tr\left[\tau^a\tau^b\tau^c\tau^ae^{i(\theta^d\tau^d-\beta)m_1}\right] \tr\left[\tau^b\tau^ce^{i(\theta^d\tau^d-\beta)m_2}\right]\right.$
        \\\quad\hfill
        $\left.+3 \tr\left[\tau^a\tau^b\tau^c\tau^ae^{i(\theta^d\tau^d-\beta)m_1}\tau^b\tau^ce^{i(\theta^d\tau^d-\beta)m_2}\right]\right\}$
    \end{tabular}
    \\\hline
    \fdiagram{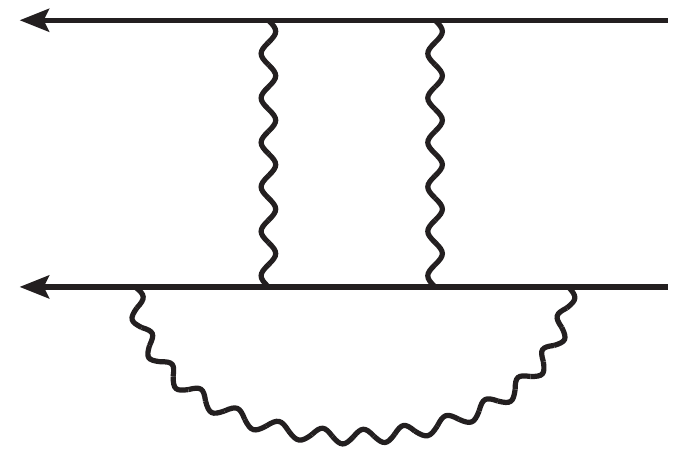}
    \quad&
    \begin{tabular}{l}
        $\frac{4}{35} \pi \left(\eta^{PS} \eta^{QR}-6 \eta^{PR} \eta^{QS}+\eta^{PQ} \eta^{RS}\right)$
        \\\quad\hfill
        $\times\left[\gamma ^N\gamma _S\gamma ^L\right]{}_{\alpha \beta } \left[\gamma ^M\gamma _P\gamma _N\gamma _Q\gamma _L\gamma _R\gamma _M\right]{}_{\gamma \delta } \left[\tau^b\tau^c\right]{}_{ij} \left[\tau^a\tau^b\tau^c\tau^a\right]{}_{kl}$,
        \\\\
        $\frac{1728}{5} \pi \left\{4 \tr\left[\tau^b\tau^ce^{i(\theta^d\tau^d-\beta)m_1}\right] \tr\left[\tau^a\tau^b\tau^c\tau^ae^{i(\theta^d\tau^d-\beta)m_2}\right]\right.$
        \\\quad\hfill
        $\left.+3 \tr\left[\tau^b\tau^ce^{i(\theta^d\tau^d-\beta)m_1}\tau^a\tau^b\tau^c\tau^ae^{i(\theta^d\tau^d-\beta)m_2}\right]\right\}$
    \end{tabular}
    \\\hline
    \fdiagram{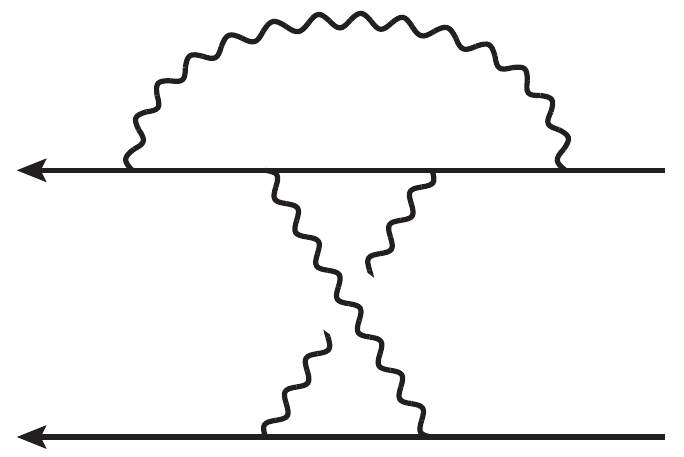}
    \quad&
    \begin{tabular}{l}
        $-\frac{4}{35} \pi \left(\eta^{PS} \eta^{QR}-6 \eta^{PR} \eta^{QS}+\eta^{PQ} \eta^{RS}\right)$
        \\\quad\hfill
        $\times\left[\gamma ^M\gamma _P\gamma ^N\gamma _Q\gamma ^L\gamma _R\gamma _M\right]{}_{\alpha \beta } \left[\gamma _L\gamma _S\gamma _N\right]{}_{\gamma \delta } \left[\tau^a\tau^b\tau^c\tau^a\right]{}_{ij} \left[\tau^c\tau^b\right]{}_{kl}$,
        \\\\
        $-\frac{1728}{5} \pi \left\{4 \tr\left[\tau^a\tau^b\tau^c\tau^ae^{i(\theta^d\tau^d-\beta)m_1}\right] \tr\left[\tau^c\tau^be^{i(\theta^d\tau^d-\beta)m_2}\right]\right.$
        \\\quad\hfill
        $\left.+3 \tr\left[\tau^a\tau^b\tau^c\tau^ae^{i(\theta^d\tau^d-\beta)m_1}\tau^c\tau^be^{i(\theta^d\tau^d-\beta)m_2}\right]\right\}$
    \end{tabular}
    \\\hline
    \fdiagram{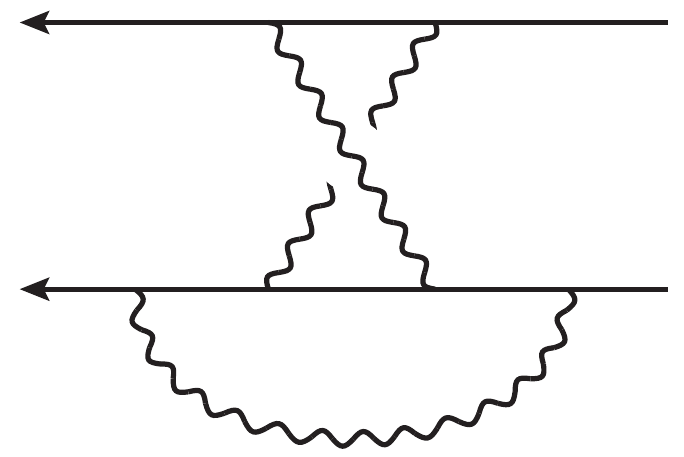}
    \quad&
    \begin{tabular}{l}
        $-\frac{4}{35} \pi \left(\eta^{PS} \eta^{QR}-6 \eta^{PR} \eta^{QS}+\eta^{PQ} \eta^{RS}\right)$
        \\\quad\hfill
        $\times\left[\gamma ^L\gamma _S\gamma ^N\right]{}_{\alpha \beta } \left[\gamma ^M\gamma _P\gamma _N\gamma _Q\gamma _L\gamma _R\gamma _M\right]{}_{\gamma \delta } \left[\tau^c\tau^b\right]{}_{ij} \left[\tau^a\tau^b\tau^c\tau^a\right]{}_{kl}$,
        \\\\
        $-\frac{1728}{5} \pi \left\{4 \tr\left[\tau^c\tau^be^{i(\theta^d\tau^d-\beta)m_1}\right] \tr\left[\tau^a\tau^b\tau^c\tau^ae^{i(\theta^d\tau^d-\beta)m_2}\right]\right.$
        \\\quad\hfill
        $\left.+3 \tr\left[\tau^c\tau^be^{i(\theta^d\tau^d-\beta)m_1}\tau^a\tau^b\tau^c\tau^ae^{i(\theta^d\tau^d-\beta)m_2}\right]\right\}$
    \end{tabular}
    \\\hline
    \fdiagram{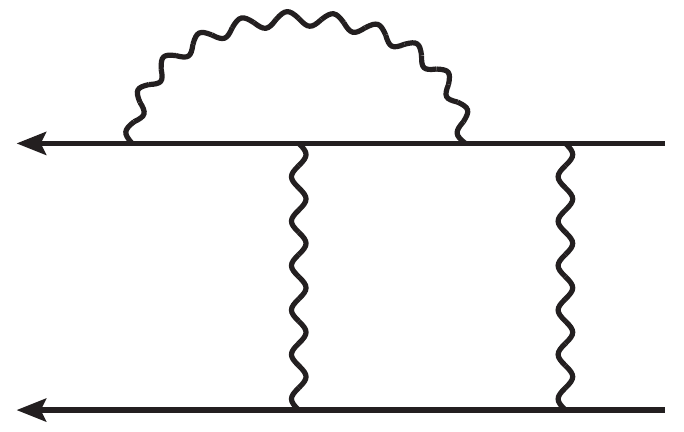}
    \quad&
    \begin{tabular}{l}
        $\frac{8}{105} \pi \left(\eta^{PS} \eta^{QR}+\eta^{PR} \eta^{QS}+8 \eta^{PQ} \eta^{RS}\right)$
        \\\quad\hfill
        $\times\left[\gamma ^M\gamma _P\gamma ^N\gamma _Q\gamma _M\gamma _R\gamma ^L\right]{}_{\alpha \beta } \left[\gamma _N\gamma _S\gamma _L\right]{}_{\gamma \delta } \left[\tau^a\tau^b\tau^a\tau^c\right]{}_{ij} \left[\tau^b\tau^c\right]{}_{kl}$,
        \\\\
        $-\frac{192}{5} \pi \left\{28 \tr\left[\tau^a\tau^b\tau^a\tau^ce^{i(\theta^d\tau^d-\beta)m_1}\right] \tr\left[\tau^b\tau^ce^{i(\theta^d\tau^d-\beta)m_2}\right]\right.$
        \\\quad\hfill
        $\left.+29 \tr\left[\tau^a\tau^b\tau^a\tau^ce^{i(\theta^d\tau^d-\beta)m_1}\tau^b\tau^ce^{i(\theta^d\tau^d-\beta)m_2}\right]\right\}$
    \end{tabular}
    \\\hline
    \fdiagram{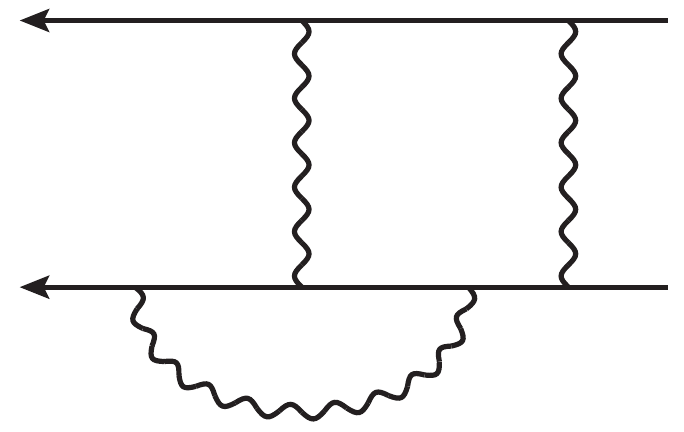}
    \quad&
    \begin{tabular}{l}
        $\frac{8}{105} \pi \left(\eta^{PS} \eta^{QR}+\eta^{PR} \eta^{QS}+8 \eta^{PQ} \eta^{RS}\right)$
        \\\quad\hfill
        $\times\left[\gamma ^N\gamma _S\gamma ^L\right]{}_{\alpha \beta } \left[\gamma ^M\gamma _P\gamma _N\gamma _Q\gamma _M\gamma _R\gamma _L\right]{}_{\gamma \delta } \left[\tau^b\tau^c\right]{}_{ij} \left[\tau^a\tau^b\tau^a\tau^c\right]{}_{kl}$,
        \\\\
        $-\frac{192}{5} \pi \left\{28 \tr\left[\tau^b\tau^ce^{i(\theta^d\tau^d-\beta)m_1}\right] \tr\left[\tau^a\tau^b\tau^a\tau^ce^{i(\theta^d\tau^d-\beta)m_2}\right]\right.$
        \\\quad\hfill
        $\left.+29 \tr\left[\tau^b\tau^ce^{i(\theta^d\tau^d-\beta)m_1}\tau^a\tau^b\tau^a\tau^ce^{i(\theta^d\tau^d-\beta)m_2}\right]\right\}$
    \end{tabular}
    \\\hline
    \fdiagram{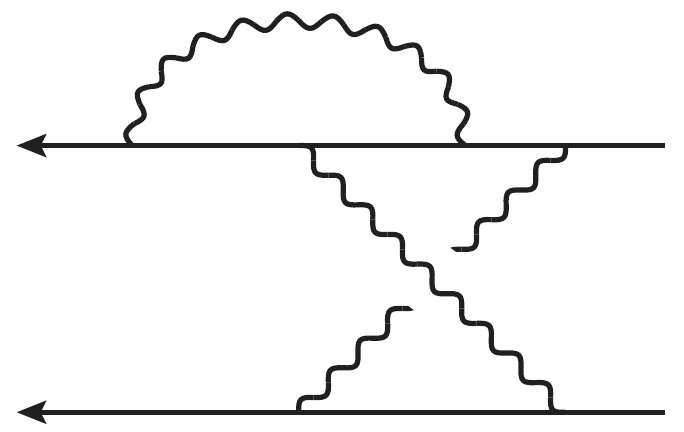}
    \quad&
    \begin{tabular}{l}
        $-\frac{8}{105} \pi \left(\eta^{PS} \eta^{QR}+\eta^{PR} \eta^{QS}+8 \eta^{PQ} \eta^{RS}\right)$
        \\\quad\hfill
        $\times\left[\gamma ^M\gamma _P\gamma ^N\gamma _Q\gamma _M\gamma _R\gamma ^L\right]{}_{\alpha \beta } \left[\gamma _L\gamma _S\gamma _N\right]{}_{\gamma \delta } \left[\tau^a\tau^b\tau^a\tau^c\right]{}_{ij} \left[\tau^c\tau^b\right]{}_{kl}$,
        \\\\
        $\frac{192}{5} \pi \left\{28 \tr\left[\tau^a\tau^b\tau^a\tau^ce^{i(\theta^d\tau^d-\beta)m_1}\right] \tr\left[\tau^c\tau^be^{i(\theta^d\tau^d-\beta)m_2}\right]\right.$
        \\\quad\hfill
        $\left.+13 \tr\left[\tau^a\tau^b\tau^a\tau^ce^{i(\theta^d\tau^d-\beta)m_1}\tau^c\tau^be^{i(\theta^d\tau^d-\beta)m_2}\right]\right\}$
    \end{tabular}
    \\\hline
    \fdiagram{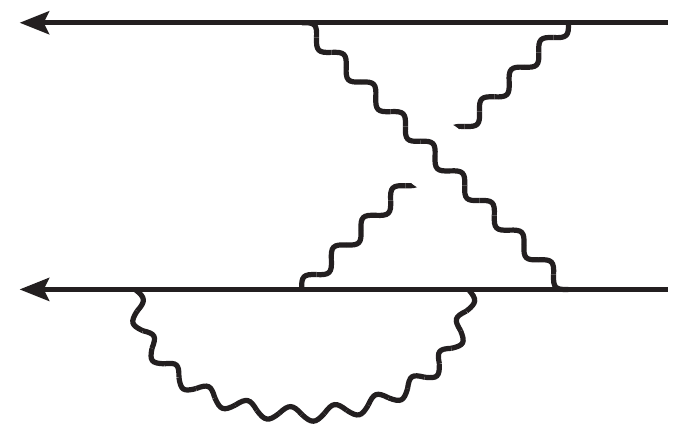}
    \quad&
    \begin{tabular}{l}
        $-\frac{8}{105} \pi \left(\eta^{PS} \eta^{QR}+\eta^{PR} \eta^{QS}+8 \eta^{PQ} \eta^{RS}\right)$
        \\\quad\hfill
        $\times\left[\gamma ^L\gamma _S\gamma ^N\right]{}_{\alpha \beta } \left[\gamma ^M\gamma _P\gamma _N\gamma _Q\gamma _M\gamma _R\gamma _L\right]{}_{\gamma \delta } \left[\tau^c\tau^b\right]{}_{ij} \left[\tau^a\tau^b\tau^a\tau^c\right]{}_{kl}$,
        \\\\
        $\frac{192}{5} \pi \left\{28 \tr\left[\tau^c\tau^be^{i(\theta^d\tau^d-\beta)m_1}\right] \tr\left[\tau^a\tau^b\tau^a\tau^ce^{i(\theta^d\tau^d-\beta)m_2}\right]\right.$
        \\\quad\hfill
        $\left.+13 \tr\left[\tau^c\tau^be^{i(\theta^d\tau^d-\beta)m_1}\tau^a\tau^b\tau^a\tau^ce^{i(\theta^d\tau^d-\beta)m_2}\right]\right\}$
    \end{tabular}
    \\\hline
    \fdiagram{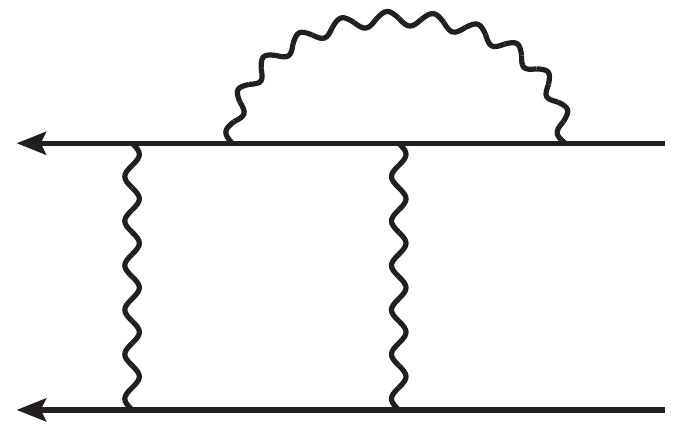}
    \quad&
    \begin{tabular}{l}
        $\frac{8}{105} \pi \left(8 \eta^{PS} \eta^{QR}+\eta^{PR} \eta^{QS}+\eta^{PQ} \eta^{RS}\right)$
        \\\quad\hfill
        $\times\left[\gamma ^M\gamma _P\gamma ^N\gamma _Q\gamma ^L\gamma _R\gamma _N\right]{}_{\alpha \beta } \left[\gamma _M\gamma _S\gamma _L\right]{}_{\gamma \delta } \left[\tau^a\tau^b\tau^c\tau^b\right]{}_{ij} \left[\tau^a\tau^c\right]{}_{kl}$,
        \\\\
        $-\frac{192}{5} \pi \left\{28 \tr\left[\tau^a\tau^b\tau^c\tau^be^{i(\theta^d\tau^d-\beta)m_1}\right] \tr\left[\tau^a\tau^ce^{i(\theta^d\tau^d-\beta)m_2}\right]\right.$
        \\\quad\hfill
        $\left.+29 \tr\left[\tau^a\tau^b\tau^c\tau^be^{i(\theta^d\tau^d-\beta)m_1}\tau^a\tau^ce^{i(\theta^d\tau^d-\beta)m_2}\right]\right\}$
    \end{tabular}
    \\\hline
    \fdiagram{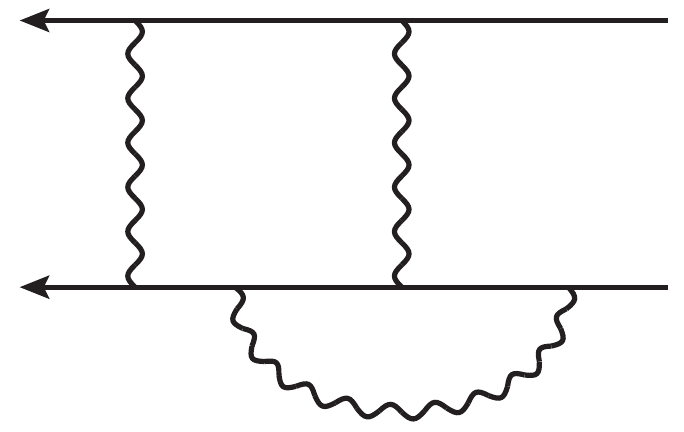}
    \quad&
    \begin{tabular}{l}
        $\frac{8}{105} \pi \left(8 \eta^{PS} \eta^{QR}+\eta^{PR} \eta^{QS}+\eta^{PQ} \eta^{RS}\right)$
        \\\quad\hfill
        $\times\left[\gamma ^M\gamma _S\gamma ^L\right]{}_{\alpha \beta } \left[\gamma _M\gamma _P\gamma ^N\gamma _Q\gamma _L\gamma _R\gamma _N\right]{}_{\gamma \delta } \left[\tau^a\tau^c\right]{}_{ij} \left[\tau^a\tau^b\tau^c\tau^b\right]{}_{kl}$,
        \\\\
        $-\frac{192}{5} \pi \left\{28 \tr\left[\tau^a\tau^ce^{i(\theta^d\tau^d-\beta)m_1}\right] \tr\left[\tau^a\tau^b\tau^c\tau^be^{i(\theta^d\tau^d-\beta)m_2}\right]\right.$
        \\\quad\hfill
        $\left.+29 \tr\left[\tau^a\tau^ce^{i(\theta^d\tau^d-\beta)m_1}\tau^a\tau^b\tau^c\tau^be^{i(\theta^d\tau^d-\beta)m_2}\right]\right\}$
    \end{tabular}
    \\\hline
    \fdiagram{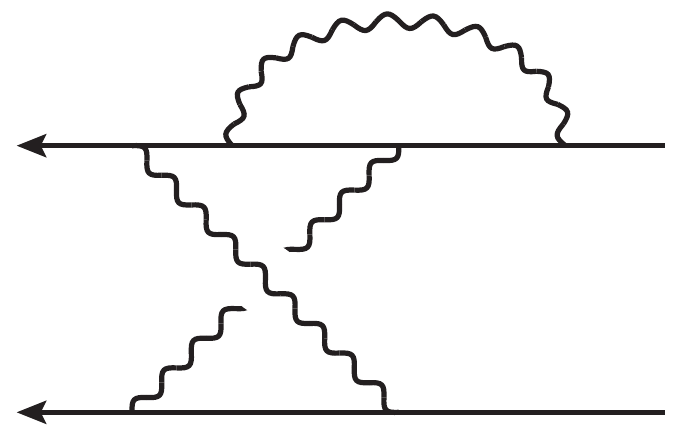}
    \quad&
    \begin{tabular}{l}
        $-\frac{8}{105} \pi \left(8 \eta^{PS} \eta^{QR}+\eta^{PR} \eta^{QS}+\eta^{PQ} \eta^{RS}\right)$
        \\\quad\hfill
        $\times\left[\gamma ^M\gamma _P\gamma ^N\gamma _Q\gamma ^L\gamma _R\gamma _N\right]{}_{\alpha \beta } \left[\gamma _L\gamma _S\gamma _M\right]{}_{\gamma \delta } \left[\tau^a\tau^b\tau^c\tau^b\right]{}_{ij} \left[\tau^c\tau^a\right]{}_{kl}$,
        \\\\
        $\frac{192}{5} \pi \left\{28 \tr\left[\tau^a\tau^b\tau^c\tau^be^{i(\theta^d\tau^d-\beta)m_1}\right] \tr\left[\tau^c\tau^ae^{i(\theta^d\tau^d-\beta)m_2}\right]\right.$
        \\\quad\hfill
        $\left.+13 \tr\left[\tau^a\tau^b\tau^c\tau^be^{i(\theta^d\tau^d-\beta)m_1}\tau^c\tau^ae^{i(\theta^d\tau^d-\beta)m_2}\right]\right\}$
    \end{tabular}
    \\\hline
    \fdiagram{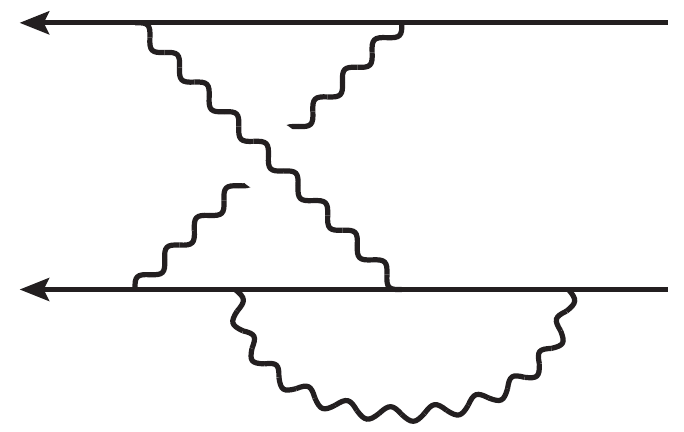}
    \quad&
    \begin{tabular}{l}
        $-\frac{8}{105} \pi \left(8 \eta^{PS} \eta^{QR}+\eta^{PR} \eta^{QS}+\eta^{PQ} \eta^{RS}\right)$
        \\\quad\hfill
        $\times\left[\gamma ^L\gamma _S\gamma ^M\right]{}_{\alpha \beta } \left[\gamma _M\gamma _P\gamma ^N\gamma _Q\gamma _L\gamma _R\gamma _N\right]{}_{\gamma \delta } \left[\tau^c\tau^a\right]{}_{ij} \left[\tau^a\tau^b\tau^c\tau^b\right]{}_{kl}$,
        \\\\
        $\frac{192}{5} \pi \left\{28 \tr\left[\tau^c\tau^ae^{i(\theta^d\tau^d-\beta)m_1}\right] \tr\left[\tau^a\tau^b\tau^c\tau^be^{i(\theta^d\tau^d-\beta)m_2}\right]\right.$
        \\\quad\hfill
        $\left.+13 \tr\left[\tau^c\tau^ae^{i(\theta^d\tau^d-\beta)m_1}\tau^a\tau^b\tau^c\tau^be^{i(\theta^d\tau^d-\beta)m_2}\right]\right\}$
    \end{tabular}
    \\\hline
    \fdiagram{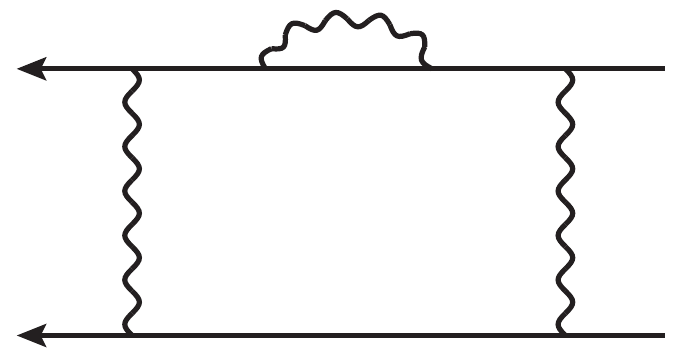}
    \quad&
    \begin{tabular}{l}
        $\frac{16}{105} \pi \left(\eta^{PS} \eta^{QR}+\eta^{PR} \eta^{QS}+\eta^{PQ} \eta^{RS}\right)$
        \\\quad\hfill
        $\times\left[\gamma ^M\gamma _P\gamma ^N\gamma _Q\gamma _N\gamma _R\gamma ^L\right]{}_{\alpha \beta } \left[\gamma _M\gamma _S\gamma _L\right]{}_{\gamma \delta } \left[\tau^a\tau^b\tau^b\tau^c\right]{}_{ij} \left[\tau^a\tau^c\right]{}_{kl}$,
        \\\\
        $\frac{64}{5} \pi \left\{52 \tr\left[\tau^a\tau^b\tau^b\tau^ce^{i(\theta^d\tau^d-\beta)m_1}\right] \tr\left[\tau^a\tau^ce^{i(\theta^d\tau^d-\beta)m_2}\right]\right.$
        \\\quad\hfill
        $\left.+51 \tr\left[\tau^a\tau^b\tau^b\tau^ce^{i(\theta^d\tau^d-\beta)m_1}\tau^a\tau^ce^{i(\theta^d\tau^d-\beta)m_2}\right]\right\}$
    \end{tabular}
    \\\hline
    \fdiagram{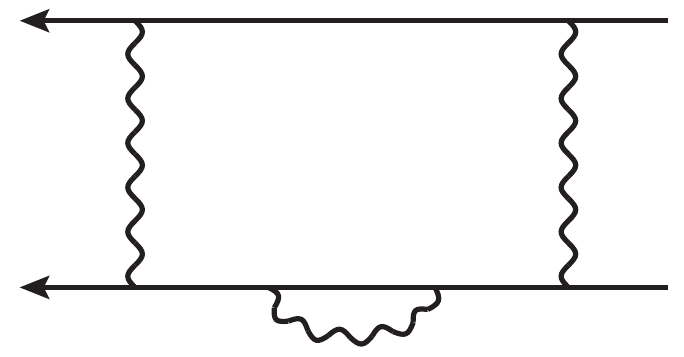}
    \quad&
    \begin{tabular}{l}
        $\frac{16}{105} \pi \left(\eta^{PS} \eta^{QR}+\eta^{PR} \eta^{QS}+\eta^{PQ} \eta^{RS}\right)$
        \\\quad\hfill
        $\times\left[\gamma ^M\gamma _S\gamma ^L\right]{}_{\alpha \beta } \left[\gamma _M\gamma _P\gamma ^N\gamma _Q\gamma _N\gamma _R\gamma _L\right]{}_{\gamma \delta } \left[\tau^a\tau^c\right]{}_{ij} \left[\tau^a\tau^b\tau^b\tau^c\right]{}_{kl}$,
        \\\\
        $\frac{64}{5} \pi \left\{52 \tr\left[\tau^a\tau^ce^{i(\theta^d\tau^d-\beta)m_1}\right] \tr\left[\tau^a\tau^b\tau^b\tau^ce^{i(\theta^d\tau^d-\beta)m_2}\right]\right.$
        \\\quad\hfill
        $\left.+51 \tr\left[\tau^a\tau^ce^{i(\theta^d\tau^d-\beta)m_1}\tau^a\tau^b\tau^b\tau^ce^{i(\theta^d\tau^d-\beta)m_2}\right]\right\}$
    \end{tabular}
    \\\hline
    \fdiagram{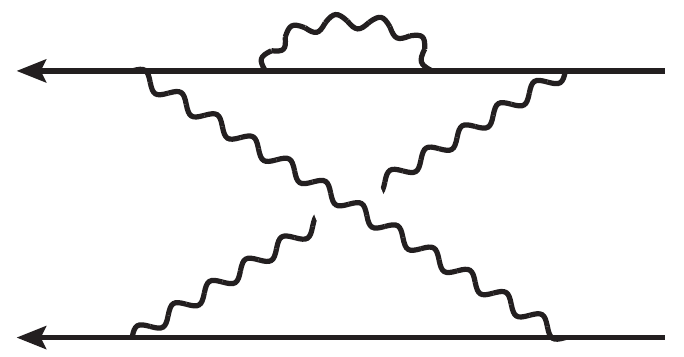}
    \quad&
    \begin{tabular}{l}
        $-\frac{16}{105} \pi \left(\eta^{PS} \eta^{QR}+\eta^{PR} \eta^{QS}+\eta^{PQ} \eta^{RS}\right)$
        \\\quad\hfill
        $\times\left[\gamma ^M\gamma _P\gamma ^N\gamma _Q\gamma _N\gamma _R\gamma ^L\right]{}_{\alpha \beta } \left[\gamma _L\gamma _S\gamma _M\right]{}_{\gamma \delta } \left[\tau^a\tau^b\tau^b\tau^c\right]{}_{ij} \left[\tau^c\tau^a\right]{}_{kl}$,
        \\\\
        $-\frac{64}{5} \pi \left\{52 \tr\left[\tau^a\tau^b\tau^b\tau^ce^{i(\theta^d\tau^d-\beta)m_1}\right] \tr\left[\tau^c\tau^ae^{i(\theta^d\tau^d-\beta)m_2}\right]\right.$
        \\\quad\hfill
        $\left.+27 \tr\left[\tau^a\tau^b\tau^b\tau^ce^{i(\theta^d\tau^d-\beta)m_1}\tau^c\tau^ae^{i(\theta^d\tau^d-\beta)m_2}\right]\right\}$
    \end{tabular}
    \\\hline
    \fdiagram{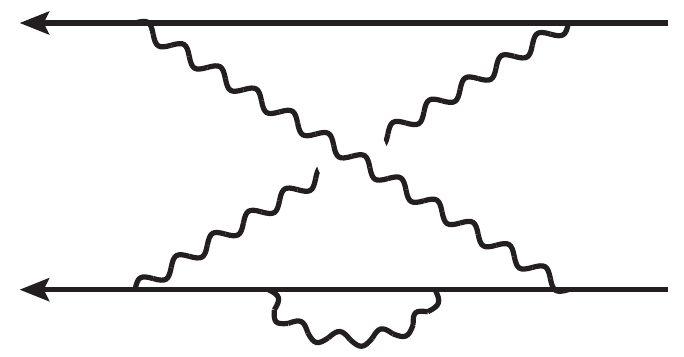}
    \quad&
    \begin{tabular}{l}
        $-\frac{16}{105} \pi \left(\eta^{PS} \eta^{QR}+\eta^{PR} \eta^{QS}+\eta^{PQ} \eta^{RS}\right)$
        \\\quad\hfill
        $\times\left[\gamma ^L\gamma _S\gamma ^M\right]{}_{\alpha \beta } \left[\gamma _M\gamma _P\gamma ^N\gamma _Q\gamma _N\gamma _R\gamma _L\right]{}_{\gamma \delta } \left[\tau^c\tau^a\right]{}_{ij} \left[\tau^a\tau^b\tau^b\tau^c\right]{}_{kl}$,
        \\\\
        $-\frac{64}{5} \pi \left\{52 \tr\left[\tau^c\tau^ae^{i(\theta^d\tau^d-\beta)m_1}\right] \tr\left[\tau^a\tau^b\tau^b\tau^ce^{i(\theta^d\tau^d-\beta)m_2}\right]\right.$
        \\\quad\hfill
        $\left.+27 \tr\left[\tau^c\tau^ae^{i(\theta^d\tau^d-\beta)m_1}\tau^a\tau^b\tau^b\tau^ce^{i(\theta^d\tau^d-\beta)m_2}\right]\right\}$
    \end{tabular}
    \\\hline
    \fdiagram{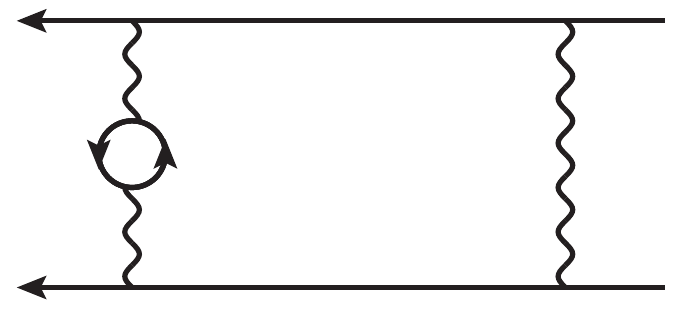}
    \quad&
    \begin{tabular}{l}
        $\frac{2}{105} \pi c_r \left(10 \eta^{PS} \eta^{QR}+3 \eta^{PR} \eta^{QS}+3 \eta^{PQ} \eta^{RS}\right)$
        \\\quad\hfill
        $\times\tr\left[\gamma ^M\gamma _Q\gamma ^L\gamma _R\right] \left[\gamma _M\gamma _P\gamma ^N\right]{}_{\alpha \beta } \left[\gamma _L\gamma _S\gamma _N\right]{}_{\gamma \delta } \left[\tau^a\tau^b\right]{}_{ij} \left[\tau^a\tau^b\right]{}_{kl}$,
        \\\\
        $\frac{256}{5} \pi c_r \left\{8 \tr\left[\tau^a\tau^be^{i(\theta^d\tau^d-\beta)m_1}\right] \tr\left[\tau^a\tau^be^{i(\theta^d\tau^d-\beta)m_2}\right]\right.$
        \\\quad\hfill
        $\left.+9 \tr\left[\tau^a\tau^be^{i(\theta^d\tau^d-\beta)m_1}\tau^a\tau^be^{i(\theta^d\tau^d-\beta)m_2}\right]\right\}$
    \end{tabular}
    \\\hline
    \fdiagram{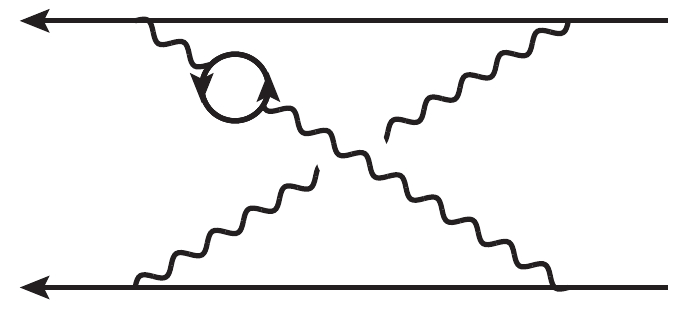}
    \quad&
    \begin{tabular}{l}
        $-\frac{2}{105} \pi c_r \left(10 \eta^{PS} \eta^{QR}+3 \eta^{PR} \eta^{QS}+3 \eta^{PQ} \eta^{RS}\right)$
        \\\quad\hfill
        $\times\tr\left[\gamma ^M\gamma _Q\gamma ^L\gamma _R\right] \left[\gamma _M\gamma _P\gamma ^N\right]{}_{\alpha \beta } \left[\gamma _N\gamma _S\gamma _L\right]{}_{\gamma \delta } \left[\tau^a\tau^b\right]{}_{ij} \left[\tau^b\tau^a\right]{}_{kl}$,
        \\\\
        $-\frac{256}{5} \pi c_r \left\{8 \tr\left[\tau^a\tau^be^{i(\theta^d\tau^d-\beta)m_1}\right] \tr\left[\tau^b\tau^ae^{i(\theta^d\tau^d-\beta)m_2}\right]\right.$
        \\\quad\hfill
        $\left.+3 \tr\left[\tau^a\tau^be^{i(\theta^d\tau^d-\beta)m_1}\tau^b\tau^ae^{i(\theta^d\tau^d-\beta)m_2}\right]\right\}$
    \end{tabular}
    \\\hline
    \fdiagram{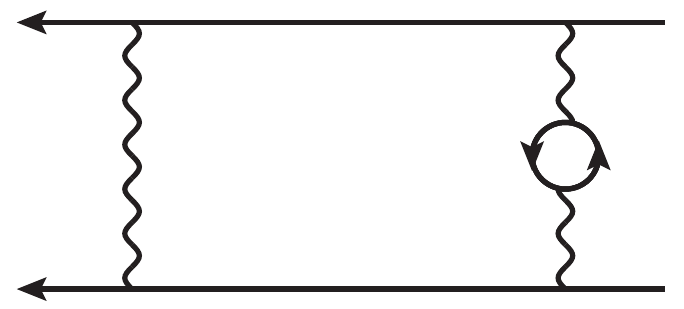}
    \quad&
    \begin{tabular}{l}
        $\frac{2}{105} \pi c_r \left(10 \eta^{PS} \eta^{QR}+3 \eta^{PR} \eta^{QS}+3 \eta^{PQ} \eta^{RS}\right)$
        \\\quad\hfill
        $\times\tr\left[\gamma ^N\gamma _Q\gamma ^L\gamma _R\right] \left[\gamma ^M\gamma _P\gamma _N\right]{}_{\alpha \beta } \left[\gamma _M\gamma _S\gamma _L\right]{}_{\gamma \delta } \left[\tau^a\tau^b\right]{}_{ij} \left[\tau^a\tau^b\right]{}_{kl}$,
        \\\\
        $\frac{256}{5} \pi c_r \left\{8 \tr\left[\tau^a\tau^be^{i(\theta^d\tau^d-\beta)m_1}\right] \tr\left[\tau^a\tau^be^{i(\theta^d\tau^d-\beta)m_2}\right]\right.$
        \\\quad\hfill
        $\left.+9 \tr\left[\tau^a\tau^be^{i(\theta^d\tau^d-\beta)m_1}\tau^a\tau^be^{i(\theta^d\tau^d-\beta)m_2}\right]\right\}$
    \end{tabular}
    \\\hline
    \fdiagram{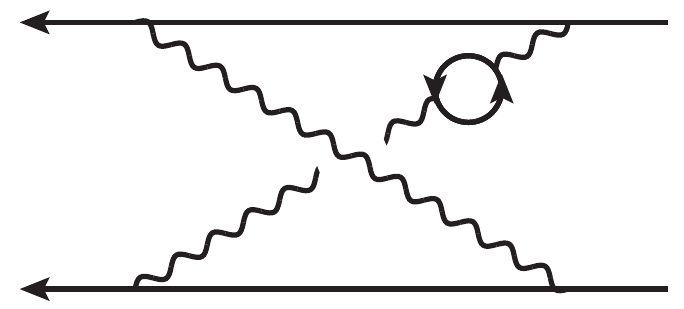}
    \quad&
    \begin{tabular}{l}
        $-\frac{2}{105} \pi c_r \left(10 \eta^{PS} \eta^{QR}+3 \eta^{PR} \eta^{QS}+3 \eta^{PQ} \eta^{RS}\right)$
        \\\quad\hfill
        $\times\tr\left[\gamma ^N\gamma _Q\gamma ^L\gamma _R\right] \left[\gamma ^M\gamma _P\gamma _N\right]{}_{\alpha \beta } \left[\gamma _L\gamma _S\gamma _M\right]{}_{\gamma \delta } \left[\tau^a\tau^b\right]{}_{ij} \left[\tau^b\tau^a\right]{}_{kl}$,
        \\\\
        $-\frac{256}{5} \pi c_r \left\{8 \tr\left[\tau^a\tau^be^{i(\theta^d\tau^d-\beta)m_1}\right] \tr\left[\tau^b\tau^ae^{i(\theta^d\tau^d-\beta)m_2}\right]\right.$
        \\\quad\hfill
        $\left.+3 \tr\left[\tau^a\tau^be^{i(\theta^d\tau^d-\beta)m_1}\tau^b\tau^ae^{i(\theta^d\tau^d-\beta)m_2}\right]\right\}$
    \end{tabular}
\end{longtable}

\bibliographystyle{apsrev4-1}
\bibliography{4F2L}
\end{document}